\documentclass[sn-nature]{sn-jnl}

\usepackage{graphicx}%
\usepackage{multirow}%
\usepackage{amsmath,amssymb,amsfonts}%
\usepackage{amsthm}%
\usepackage{mathrsfs}%
\DeclareFontFamily{U}{rsfs}{\skewchar\font127 }
\DeclareFontShape{U}{rsfs}{m}{n}{%
   <-6> rsfs5
   <6-8> rsfs7
   <8-> rsfs10
}{}
\usepackage[title]{appendix}%
\usepackage{xcolor}%
\usepackage{textcomp}%
\usepackage{manyfoot}%
\usepackage{booktabs}%
\usepackage{algorithm}%
\usepackage{algorithmicx}%
\usepackage{algpseudocode}%
\usepackage{listings}%
\usepackage{natbib}

\theoremstyle{thmstyleone}%
\theoremstyle{thmstyletwo}%

\theoremstyle{thmstylethree}%

\begin{document}

\title{What Types of Human-AI Teams Exist?}


\author*[1]{\fnm{Nathan} \sur{Hughes}}\email{nathan.hughes@york.ac.uk}

\author[1]{\fnm{Ibrahim} \sur{Habli}}\email{ibrahim.habli@york.ac.uk}

\affil*[1]{\orgdiv{Centre for Assuring Autonomy}, \orgname{University of York}, \orgaddress{\street{Deramore Lane}, \city{York}, \state{North Yorkshire}, \postcode{YO10 5DD}, \country{UK}}}

\abstract{Human-AI teaming has received increasing attention in the literature. However, the range of studies conducted in multiple domains make it difficult to understand what types of teams are being studied, and in what ways are they similar/different from one another. In this study, we analyse 53 papers on human-AI teams and categorise them into five main clusters based on psychological taxonomies of teaming; AI Assistant, Ad-hoc Dependency, Ad-hoc Forced Dependency, Paired Equanimity, and Group Equanimity. Each cluster represents a unique combination of holistic team-level characteristics, indicating there are multiple disparate team types studied under the same definition. In turn, this raises the question of whether insights are truly transferable between papers. We conclude with guidance on how to identify the types of human-AI teams studied, a checklist for reporting a human-AI team in research work, and ways in which the field can be further synthesised.}




\maketitle

\section{Introduction}\label{int}

Human-AI teaming as a field has grown exponentially popular in recent years, and broadly refers to a team consisting of one or more humans and one or more AI working together interdependently towards a shared goal \cite{berretta2023defining}.
To do so, each entity must act as a team member, and possess unique and complementary capabilities.
In other words, human-AI teaming aims to capitalise on the complementary strengths of both humans and AI, with the goal of improving overall team performance, across several applications.
For example, it has been studied in numerous domains, ranging from safety-critical systems such as healthcare (e.g. \cite{bienefeld2023human}), work-based settings (e.g. \cite{lu2023readingquizmaker}), and recreational settings such as video games (e.g. \cite{flathmann2023examining}).

Given this explosion of interest, there have been recent reviews on the human-AI teaming literature, such as \cite{berretta2023defining}.
However, such reviews reveal many human-AI teaming papers focus heavily on the `AI' technical aspect, rather than on the 'team' aspect.
This represents an important gap in understanding, as teaming is the concept that separates human-AI teaming from other related terminologies, such as human-AI collaboration \cite{wang2020human} and AI decision-support tools \cite{vcartolovni2022ethical}.
Further, knowing who is involved in a team is only one component of what distinguishes and explains a team \cite{wildman2012task}, and in particular overlooks which holistic characteristics explain currently studied human-AI teams.

This leads to two problems.
Firstly, it is difficult to extract concrete examples of when an AI is no longer a `mere tool' and instead perceived as a teammate from existing reviews.
In turn, the specificity of the term is difficult to explain and use to separate different human and AI contexts.
Secondly, and more crucially, it is not clear what types of teams are studied within human-AI teaming from a teaming perspective.
In particular, it is unclear what types of tasks are pursued by humans and AI, and in what ways are team members organised to allow for collaborative working.
This is important to consider, as the wide range of application domains referenced earlier makes it unclear what these disparate environments have in common, if anything.
Overall, it has become unclear what is specific about human-AI \textit{teaming}, and in turn what specifically has been studied under this umbrella definition.
In turn, it is difficult to understand what is cohesive about this line of research, and how effectively insights can be shared across papers.

To resolve these issues in definitional uncertainty, recent reviews have called for more research from a teaming perspective, rather than one that solely focuses on the AI involved \cite{berretta2023defining}.
Analysing existing papers through such a teaming perspective, and consequently an interdisciplinary lens, is an initial step in this process.
Furthermore, as existing definitions for human-AI teams were originally inspired from psychological theories on teaming (e.g. \cite{mcneese2018teaming}), a natural starting point for understanding human-AI teams from a teaming perspective is to apply psychological taxonomies of teams to human-AI teams.
In doing so, we may better understand and categorise what kinds of teams are studied within human-AI teaming, as well as understand how human-AI teams are unique from all-human teams.

In this paper, we apply psychological taxonomies of teaming to understand studies conducted within human-AI teaming research.
A scoping review revealed 53 experimental papers on human-AI teaming, which were categorised based on the taxonomies provided by \cite{wildman2012task} via a deductive content analysis \cite{krippendorff2018content}.
Five main types of team were found, each distinct and with unique combinations of team level characteristics.
Consequently, whilst there is a large body of work on human-AI teaming, the teams studied are not inherently interchangeable, despite sharing the same overarching definition.
In turn, the ability to synthesise and confidently transfer key insights between similar studies is reduced.
The findings also reveal differing implicit assumptions between papers on what makes a human-AI team a `team,' which requires further clarification.

There are therefore three main contributions of this work.
Firstly, we contribute an initial taxonomy that classifies types of human-AI teams studied in the literature.
Secondly, we contribute a discussion on issues arising from the current approach to classifying all studies under the same definition of human-AI teaming, despite this covering a wide and disparate number of teaming types.
Finally, we conclude the work with guidance on how to report human-AI teams in future papers to aid in clarity and specificity.
The checklist provided aims to inspire reflection on what types of teams we as a field are interested in pursuing, as well as more concrete reasons as to why.

\section{Background}

\subsection{Human-AI teaming}

Several papers have, as part of their work, offered a definition of human-AI teaming.
For example, \cite{zhang2021ideal} defines them as ``a mixed entity of two or more subjects (i.e. at least one each being human or AI), who interact interdependently and perform shared tasks to achieve the same valued goals.''
From a recent literature review and synthesis of literature, a more comprehensive definition has been proposed by \cite{berretta2023defining}.
Bolded text emphasises the key concepts underpinning the definition:

\begin{quote}
    ``Human-AI teaming is a \textbf{process} between one or more human(s) and one or more (partially) autonomous AI system(s) \textbf{acting as team members} with \textbf{unique and complementary capabilities}, who work \textbf{interdependently} toward a \textbf{common goal}. The team members’ roles are \textbf{dynamically adapting} throughout the collaboration, requiring \textbf{coordination and mutual communication} to meet each other’s and the task’s requirements. For this, a \textbf{mutual sharing of intents}, shared \textbf{situational awareness} and developing shared \textbf{mental models} are necessary, as well as \textbf{trust} within the team.''
\end{quote}

As can be seen from these definitions, despite being composed of multiple concepts, they remain abstract and high level in nature, leaving room for a wide range of researcher interpretation.
In other words, it is unclear how practically useful such definitions are for a researcher interested in identifying and designing a human-AI team.
For example, how would one use `acting as team members with unique and complementary capabilities' to distinguish an AI teammate from a mere AI tool?
What does it mean to share a goal with an AI?
Is the type of common goal important, and at what level of abstraction should the goal be considered (e.g. task level, team level)?
Overall, there is not a clear sense for where the `team' component exists in these definitions, outside of a `perception' that an AI is a teammate.
Consequently, how such perceptions would manifest, or be measured reliably at face value, is unclear.

It is therefore unsurprising that literature reviews have highlighted many papers do not have a clear foundation of what the teaming concept refers to, and it was very rare to see connections between psychological literature on teaming and human-AI teaming papers \cite{berretta2023defining}.
This is interesting, as human teams are the inspiration behind the definition itself; O'Neill et al., 2022 \cite{o2022human} for example is commonly referenced as a definitional paper on human-AI teaming (despite itself referring to human-\textit{autonomy} teaming), which links its conception of teams to \cite{mcneese2018teaming}.
In turn, this paper sources its definition directly from psychological literature on teaming \cite{salas2008teams}.

This ambiguity has led to a very wide range of the types of teams being studied, all under the same definition.
For example, \cite{smith2025navigating} considers how a human and an AI would work together to allow permitted personnel into a restricted building via the facial recognition. 
Conversely, \cite{duan2024understanding} considers how two human teammates would work with an AI to perform reconnaissance photography as part of an aerial-based mission.
Both studies refer to these setups as human-AI teams, however their similarities outside of the presence of an AI are limited.
The level of role adaption, number of teammates, unique capabilities each possess, and the extent to which an AI is acting as a teammate for example, varies significantly between settings.

Overall, it is difficult to understand what a human-AI team would look like from reading the overarching definitions, and given the wide range of applications studied, looking at any individual studies would likely be similarly confusing.
There is therefore a need to understand what exactly is being studied and identify types of teams studied, from a teaming perspective.
Indeed, \cite{berretta2023defining} explicitly calls in their discussion for a ``holistic approach involving multiple disciplines,'' and for research to bring ``the teaming idea, and established theories and empirical research from human-human teaming, into the field.''
Consequently, in this work we apply existing psychological taxonomies of teams to human-AI teams, to better understand what types of human-AI teams are currently studied.

\subsection{Teaming taxonomies}
\label{sec:team}

Within the psychological literature on teaming there have been several taxonomies over the years, creating a large yet overlapping field to draw insights from.
To address this, \cite{wildman2012task} provided a synthesis of the main taxonomies into a combined model, which accounts for both task level (i.e. what actions the team performs) and team level characteristics (i.e. the overall composition of the team).
We use the synthesis taxonomies provided by \cite{wildman2012task} to identify different types of teams studied within human-AI teaming, for two further reasons.
Firstly, the team level characteristics were designed to be discrete categories, allowing for easy clustering of differences between papers.
Secondly, the taxonomies were created by the same authors that inspired the original human-AI teaming definition (i.e. human autonomy teaming, where \cite{mcneese2018teaming} references \cite{salas2008teams}).
As such, they offer an applicable lens to view human-AI teams, with discrete categories allowing for easy creation of teaming types.
The taxonomies are overviewed here, and linked to previous human-AI teaming literature where relevant.

\subsubsection{Task level characteristics}

Task level characteristics are the types of work a team member can engage in, and were specifically designed in this taxonomy to be mutually exclusive and exhaustive categories.
As explained by \cite{wildman2012task} (Pages 107-111), the following task level characteristics can be performed by members of a team:

\begin{itemize}
    \item \textbf{Managing others}: Directing, supervising, or overseeing the work of others in an authoritative role. Intends to encourage productivity amongst subordinates. Does not include managing processes, which are a type of problem-solving
    \item \textbf{Advising others}: Providing consultative professional support (e.g. expert assistance or advice) where the advisor lacks authority over the advisee
    \item \textbf{Human Service}: Providing a good or service to another party via social interaction. Intends to bring satisfaction to a client/customer
    \item \textbf{Negotiation}: Two or more parties in conflict seeking to resolve differences and reach agreement via social interaction. Does not include collaborative efforts to reach a common objective/goal, which is a type of problem-solving
    \item \textbf{Psychomotor action}: Technical and/or motor functioning requiring psychological processing to perform calculated or elaborate movements (e.g. manipulation/use of a product/machine, or a task achieved by engaging in psychomotor action)
    \item \textbf{Defined problem-solving}: Problem solving tasks with predetermined or conclusive solutions or correct answers (e.g. yes or no, option A, B, or C). Involves choosing between two or more options rather than generating a new, unique solution
    \item \textbf{Ill-defined problem-solving}: Problem solving tasks lacking predetermined or conclusive solutions or correct answers (e.g. planning, knowledge generation)
\end{itemize}

Each task type can be completed by one or more members of a team, and a team member may engage in multiple task types within a given teaming situation.
For example, a team mate may be in charge of classifying an object (defined problem-solving), as well as providing a suggestion for an action based on this classification to another teammate who is in charge of the task (advising others).
Therefore, it is possible for two teams to have the same types of tasks as part of the work, whilst on the surface appearing very different.
For example, working with an AI teammate to score goals in a video game has the same task types as working with an AI to pilot an aircraft (defined problem-solving, ill-defined problem-solving, and psychomotor action).
The inverse is also true; two teams may appear similar in setup, but in fact require very different tasks.
For example, as part of disaster response simulations it is common to navigate difficult terrain to locate and rescue victims.
However, depending on the team setup this will affect the specific tasks involved.
For example, some rescue operations will involve team members to physically rescue victims (i.e. psychomotor action), however some teams may be purely strategic in allocation of rescue attempts, thereby being exclusive to defined and ill-defined problem-solving.

Consequently, whilst task level analysis is useful to describe and understand what types of work are engaged in and by whom, they are not useful for overall team classifications.
Instead, team-level characteristics can be considered, explained in the following subsection.

\subsubsection{Team-level characteristics}

Team-level characteristics present a holistic understanding of a team, used to describe a team at a specific moment in time, as the makeup and structure may naturally shift or change during its life.
As explained by \cite{wildman2012task} (Pages 115-119), teams can be categorised using the following variables, each with their own discrete categories.

\textbf{Task interdependence} is how much the outcomes of team members are influenced by/depend on the actions of others.
It has been noted as an important factor in Human-AI Teaming (e.g. \cite{gao2023agent, flathmann2024empirically}), and one that separates it from simple tool use (e.g. \cite{zhang2021ideal, hemmer2025complementarity}).
There are four types: \textit{Pooled}, where each team member contributes to the outcome without interacting with other group members; \textit{Sequential}, where one team member must act before another can act; \textit{Reciprocal}, a one-on-one format of back-and-forth style interaction between team members (but not multiple members at once); and \textit{intensive}, where all team members interact as a unit to jointly collaborate.

\textbf{Role structure} is the extent to which roles are fundamentally different/not interchangeable versus each person is capable of performing every component.
This is not referring to team members who simply perform different roles; the key difference is whether all team members \textit{could} perform all roles.
This may not always be true, such as in the case of a role requiring specific skills or knowledge, such as a software engineer in a team of researchers.
In the case of human-AI teaming, humans are typically trained in the task at hand, and therefore have expertise/specialisation an AI does not have access to.
There are consequently two types: \textit{Functional}, where team members perform fundamentally different roles due to their level of expertise/specialisation; and \textit{divisional}, where team members perform a specific part of the overall task, but are capable of performing each part.

\textbf{Leadership structure} refers to the pattern/distribution of leadership functions, such as choosing a direction and aligning goals among team members.
There are four types: \textit{External manager}, where leadership roles are performed by someone outside of the team; \textit{Designated}, where leadership roles are performed by one member of the team, across time and tasks; \textit{Temporary}, where leadership roles are rotated across members of the team, but not held at the same time; and \textit{distributed}, where leadership roles are performed by multiple members of the team simultaneously.

\textbf{Communication structure} is the pattern/flow of communication and information sharing among team members, both verbal and non-verbal.
It has also been noted as an important factor in human-AI teaming (e.g. \cite{cabour2023explanation, zhang2023investigating}), as how humans and AI communicate can impact the effectiveness of the team's performance.
There are three types: \textit{Hub-and-wheel}, where communication passes through a central team member, often but not always the team leader, before dissemination to all other team members; \textit{Chain}, where communication passes `up and down' based on a hierarchical structure, such as rank or leadership position; and \textit{star}, where communication passes freely to and from all team members with no central point of contact or hierarchical structure.

\textbf{Physical distribution} is the spatial location of team members in relation to one another.
Whilst recent attention has been on the increased use of virtual teams aided by technology, this is separate than AI performing within the team itself.
There are three types: \textit{Colocated}, where team members are close enough to easily communicate face-to-face; \textit{Distributed}, where team members are far enough that most communication is computer-mediated (e.g. e-mail or video); and \textit{mixed}, where some team members are colocated and some are distributed.

Finally, \textbf{team life span} is the length of time a team exists as a functional, active unit.
This is not the same as the time taken to perform a complete `cycle' of the main task of the team, but rather how long the team is functional.
As such, how often the team meets and who is in the team may change over time, but this does not affect the team lifespan itself.
Life span is one way that human-AI teaming is different than simple AI tool use, as teams are typically considered to be intentionally formed (e.g. \cite{loper2023evolving, attig2024more}) and exist for more than one interaction (e.g. \cite{mcgrath2025collaborative, flathmann2021modeling}).
There are two types: \textit{Ad hoc}, where teams brought together to address a specific event before disbanding (e.g. emergency response team); and \textit{long term}, where teams exist for an extended period of time (e.g. a management team that exists for as long as its organisation exists).

The characteristics can be considered in isolation, or combined to create a holistic description of the team.
For example, a team could have intensive task interdependence, divisional roles, designated leadership structure, hub-and-wheel communication, mixed physical distribution, and an ad hoc lifespan.
This may describe a team of volunteers and medics assembled to rescue victims from a natural disaster, where there is a central communicator coordinating different members.

Overall, the team level characteristics outlined here are designed to explain how teams operate as a whole, rather than what team members perform together and individually.
This holistic account better highlights what is unique about a given team, and is consequently useful for considering how Human-AI Teams behave a unit.
However, given the nature of AI and its increasing role in teaming environments, it is also useful to understand what types of work humans and AI are being asked to perform. 
Therefore, in this paper, we apply the above taxonomies to experimental studies of Human-AI Teaming, to reveal how teaming is currently conceptualised.

\section{Methods}\label{sec11}

We performed a scoping review of experimental studies on Human-AI Teaming, in order to classify current work into the taxonomies by \cite{wildman2012task} described above.

\subsection{Paper screening}

A scoping review was conducted to collect experimental studies on Human-AI Teaming, as these reviews are useful for clarifying concepts and examining how research is being conducted in a field \cite{munn2018systematic}.
For reporting the methodology and results, we followed the PRISMA extension guidance for scoping reviews \cite{tricco2018prisma}.

\begin{figure}[h]
    \centering
    \includegraphics[width=0.7\linewidth]{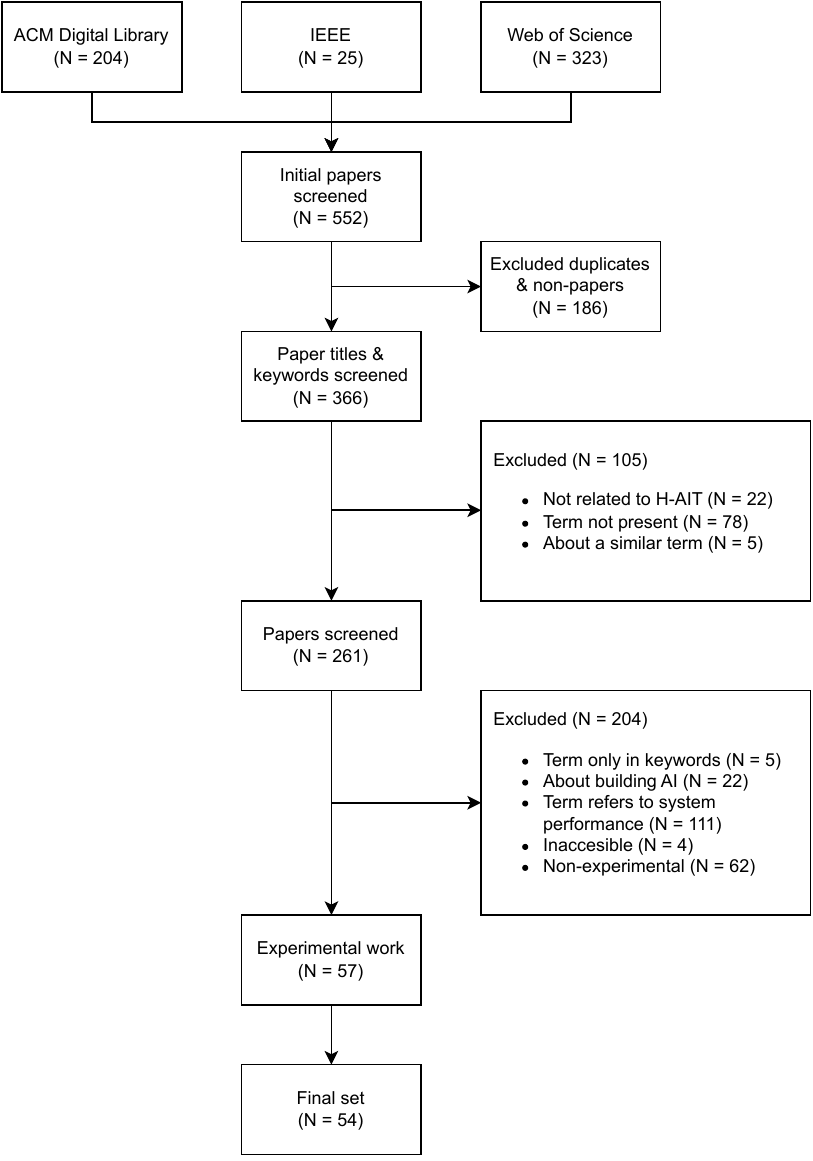}
    \caption{Paper screening process following the PRISMA method.}
    \label{fig:PRISMA}
\end{figure}

The steps taken to identify papers eligible for inclusion are shown in Figure \ref{fig:PRISMA}.

To find the initial papers, the ACM Digital Library, IEEE, and Web of Science were searched.
Eligible papers were those dating anytime before April 2025 and written in English.
The search term [``human ai team*''] was used, and papers specifically discussing Human-AI teaming were collected, and screened for those that describe experimental studies.

The screening process was conducted by the first author, after deliberation with the co-author to create the inclusion/exclusion criteria.
The initial search resulted in 204 papers from ACM, 25 from IEEE, and 323 from Web of Science, for a total of 552 papers.
After removing conference proceeding summary documents, citations, books, and duplicates, this was reduced to 366.
The titles of the papers and any keywords were then screened to find those specific to Human-AI Teaming (i.e. containing the phrase 'human-AI team*').
Papers that did not use human-AI teaming in the title or keywords were excluded, leaving 288 papers.
Papers unrelated to Human-AI Teaming were removed, as well as papers on a similar yet distinct topic (e.g. human-AI collaboration), leaving 261 papers.

These papers were then read by the first author to assess if they studied Human-AI Teaming.
Five papers only used the phrase in the keyword and nowhere else in the text, and four papers were inaccessible, and so were removed, leaving 252 papers.
Many papers used the phrase human-AI teaming to refer only to the performance of a system that included humans and AI, and provided no further information about the setup.
These papers were consequently removed, leaving 141 papers.
Finally, papers that were specific to building AI systems were removed, leaving 119 papers.

Of these papers, 57 involved an experimental design.
However, four involved no AI in the experimental design (Papers 146, 151, 166 and 176), and one paper involved two studies in the same paper (Paper \cite{hemmer2025complementarity}).
Therefore, the analysed dataset included 53 papers, and led to the analysis of 54 studies.

\subsection{Data Analysis}

Once papers were collected, the next step was to broadly map the relevant literature by applying qualitative analysis \cite{arksey2005scoping}.
As we were interested in going beyond a narrative description of experimental studies, a deductive content analysis was selected for paper analysis \cite{krippendorff2018content}.
This means applying an existing codebook to a dataset.
We used the task-level and team-level characteristic taxonomies presented by \cite{wildman2012task} (described in Section \ref{sec:team} previously) to code the types of experimental studies present in the field.

To do so, information including the number of humans and AI present, the tasks and roles assigned to humans and AI, and the team goal, were extracted.
In other words, information about what humans and AI were asked to do, in pursuit of what goal, within the human-AI team, was extracted for each paper.
Further, to aid in describing the studies, descriptives such as intended application domain and experimental environment were also extracted.

As the taxonomies were built for human only teams, it was necessary to make some minor reinterpretations to apply the team-level characteristics to AI.
The task-level characteristics mapped well to the types of tasks performed by humans and AI, apart from psychomotor action.
As interacting with an AI involves interacting with some form of computer, this could be interpreted as always involving technical functioning to achieve an outcome.
This would reduce the usefulness of the category, and so we instead chose to consider psychomotor actions only when a team member required interaction with something other than the computer used to communicate with an AI (e.g. controlling an aircraft, moving an avatar around a map).
However, the team-level characteristics of physical distribution and team-life span proved more challenging to apply.
As an AI is usually situated within a computer, it was difficult to interpret the physical distribution of the team.
Similarly, not all papers specified how long the team should typically last for, as there were many experimental designs.
Therefore, physical distribution was interpreted to mean whether the human and AI were in the same room/environment, and team life span was interpreted as whether the task at hand, if conducted in a real world scenario, was likely to be performed once (i.e. ad hoc), or part of a longer term working arrangement (i.e. long term).

Using the task-level and team-level characteristic taxonomies, a deductive content analysis was performed by the first author.
This involved assigning any relevant task-level characteristic to the extracted human and AI roles, and one discrete label from each category of the team level characteristics to summarise team being studied.
Once complete, clusters of types of human teams were created by using the team-level characteristics and isolating those with the same configurations.
Team-level rather than task-level characteristics were used as the team level contained mutually exclusive, discrete labels, allowing for easy separation of studies.
However, task level characteristics of humans and AI in each cluster were extracted in order to aid in describing each cluster, along with the typical types of task, application domains, and number of human and AI members in the team.
Further, as the categories of physical distribution and team life span proved difficult to interpret in the context of human-AI teaming, and were reinterpreted from their original meaning, these were not used to cluster the studies.
However, in clusters where there was a high percentage of a certain type of these two categories, it is mentioned for purposes of description only.

In total, seven clusters with at least two studies were identified, of which five had at least three studies, and three had at least six.
Consequently, eight studies did not form a cluster, meaning they represented unique combinations of team level characteristics.
For brevity, we only report clusters with at least three studies, however the full analysis can be found in the appendix.

\section{Results}\label{sec2}

\subsection{Sampled Papers Overview}

\begin{figure}[h]
    \centering
    \includegraphics[width=0.8\linewidth]{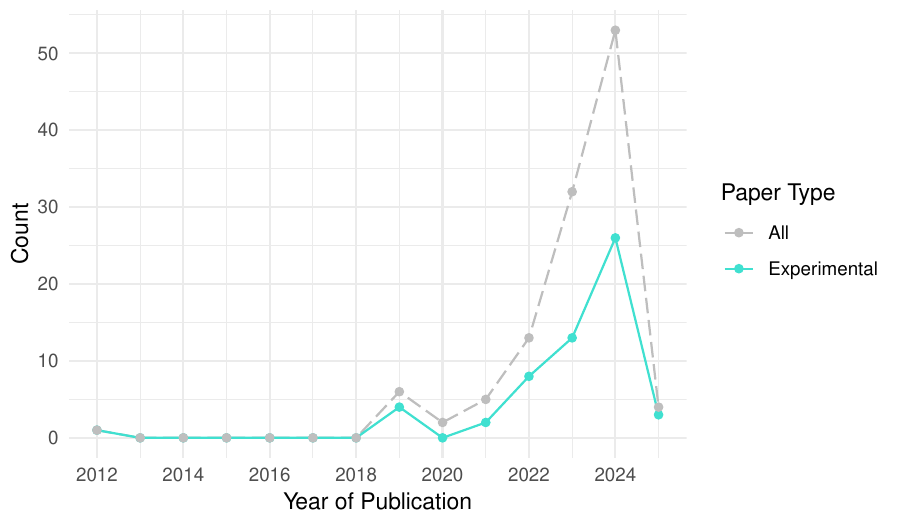}
    \caption{The publication frequency of papers that use the term Human-AI Teaming in the title or keywords. The blue line represents experimental studies, of which 54/57 are analysed in this paper.}
    \label{fig:pub}
\end{figure}

In total, there were 54 experimental studies on Human-AI Teaming in the literature, of which the majority were published in the last two years.
The publication rate is shown in Figure \ref{fig:pub}.

\begin{figure}[h]
    \centering
    \includegraphics[width=0.8\linewidth]{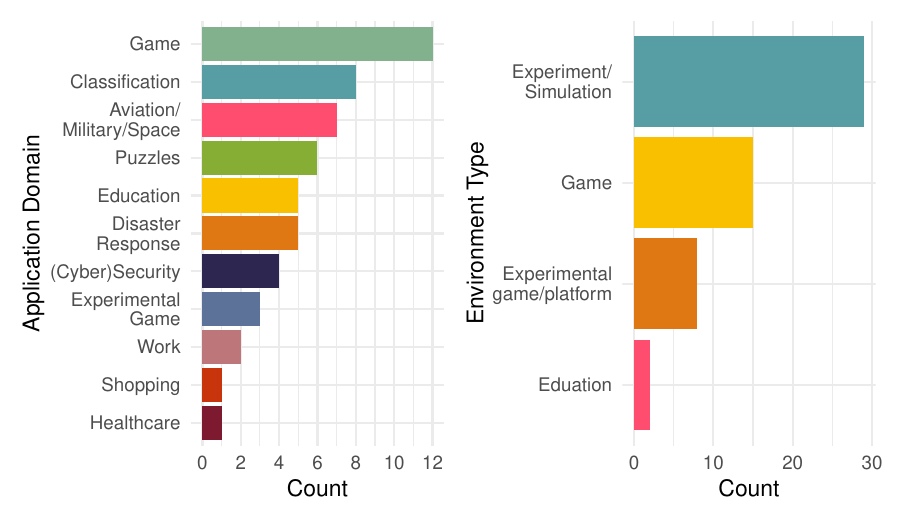}
    \caption{The type of application domain (left) and type of experimental environment (right) for the experimental studies.}
    \label{fig:appex}
\end{figure}

\begin{figure}[h]
    \centering
    \includegraphics[width=0.8\linewidth]{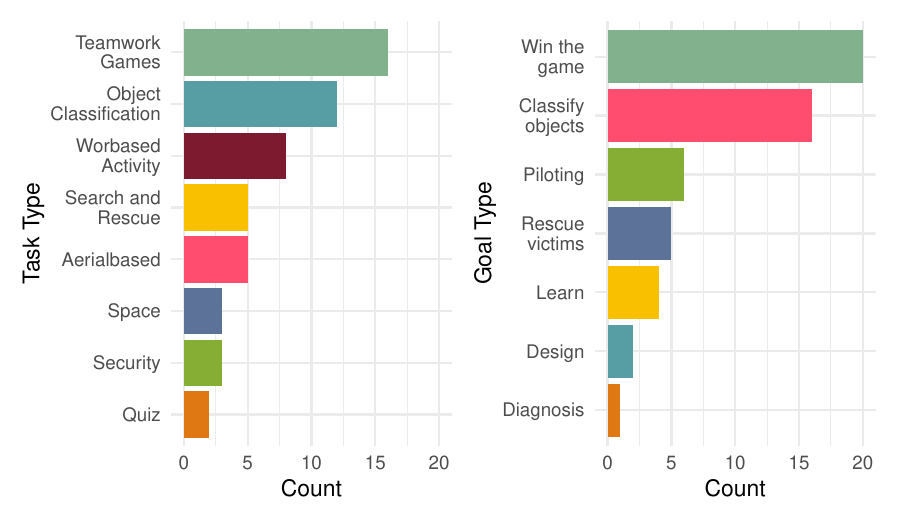}
    \caption{The type of tasks (left) and type of team goals (right) present in the experimental studies.}
    \label{fig:goaltask}
\end{figure}

\begin{figure}[h]
    \centering
    \includegraphics[width=0.8\linewidth]{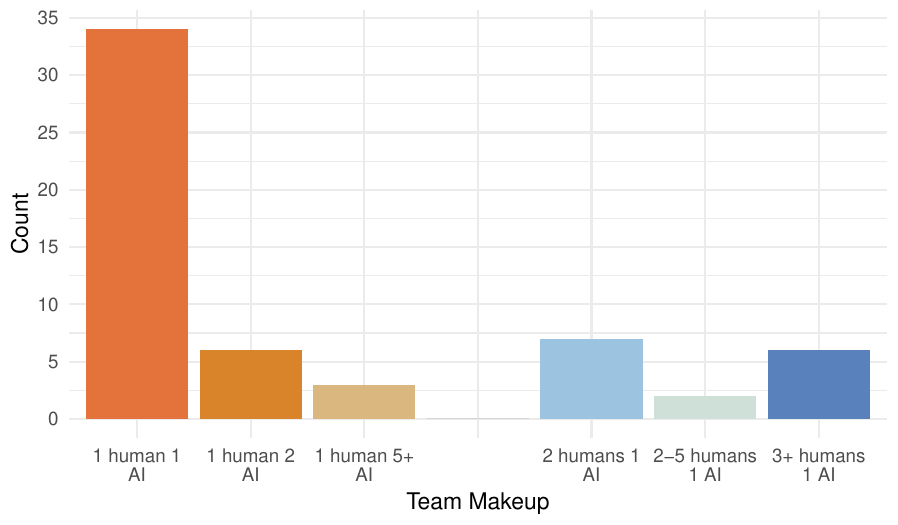}
    \caption{The number of humans and AI present in the experimental studies. Note that four papers contained experimental conditions with 1 human 2 AI as well as 2 humans 1 AI; these were split to create an overall number of 58.}
    \label{fig:makeup}
\end{figure}

An overview of the experimental papers is provided in Table \ref{tab:over}.
Figures \ref{fig:appex}, \ref{fig:goaltask} \& \ref{fig:makeup} highlight the trends in application domains, study environments, team makeups, and task types.

\begin{sidewaystable}[!]
\resizebox{\textwidth}{!}{
\begin{tabular}{@{}lllllll@{}}
\toprule
\textbf{Paper} & \textbf{Application Type} & \textbf{Team Makeup} & \textbf{Environment} & \textbf{Goal} & \textbf{Task} & \textbf{Task Type} \\ \midrule
\cite{bansal2019case} & Experimental Game & 1 human 1 AI & CAJA & identify defective vs non-defective boxes & classifying objects & Object Classification \\
\cite{olla2024cybersecurity} & Cybersecurity/Security & 1 human 1 AI & Cybersecurity game & detect cybersecurity threats & detect cybersecurity threats & Security \\
\cite{amresh2023minecraft} & Disaster Response & 3 humans 1 AI & Minecraft & rescue survivors & rescue survivors & Search and Rescue \\
\cite{schwalb2022study} & Disaster Response & 1 human 1 AI & Simulation & rescue hostages & rescue hostages & Search and Rescue \\
\cite{tariq2025a2c} & Cybersecurity/Security & 1 human 1 AI & Simulation & detect cybersecurity threats & detect cybersecurity threats & Security \\
\cite{josephs2023artifact} & Cybersecurity/Security & 1 human 1 AI & Simulation & detect deepfakes & deepfake detection & Object Classification \\
\cite{ong2012closing} & Game & 1 human 1 AI & Defend the Pass & win the game & kill all monsters game & Teamwork Games \\
\cite{zvelebilova2024collective} & Puzzles & 3-4 humans 1 AI & Puzzle & win the game & solve pattern recognition puzzle & Teamwork Games \\
\cite{xu2023comparing} & Classification & 1 human 1 AI & Experiment & annotate faces & face detection & Object Classification \\
\cite{hemmer2025complementarity} & Classification & 1 human 1 AI & Experiment & classify objects & appraise house prices & Object Classification \\
\cite{hemmer2025complementarity} & Classification & 1 human 1 AI & Experiment & classify objects & image classification & Object Classification \\
\cite{zhang2024crew} & Experimental Game & 1 human 1 AI & Experimental games & win the game & bowling; maze; hide and seek & Teamwork Games \\
\cite{munyaka2023decision} & Game & 1 human 2 AI & Guess the Word & win the game & guess the word game & Teamwork Games \\
\cite{mahmood2024designing} & Classification & 1 human 1 AI & Experiment & classify objects & classifying objects & Object Classification \\
\cite{newn2019designing} & Game & 1 human 1 AI & Ticket to Ride & win the game & connect cities game & Teamwork Games \\
\cite{bansal2021does} & Classification & 1 human 1 AI & Experiment & classify sentiment of reviews & sentiment analysis & Object Classification \\
\cite{flathmann2024empirically} & Game & 1 human 1 AI & Rocket League & score the most goals & score in rocket league & Teamwork Games \\
\cite{flathmann2023examining} & Game & 1 human 1 AI & Rocket League & score the most goals & score in rocket league & Teamwork Games \\
\cite{jorge2024should} & Shopping & 1 human 2 AI & Simulation & fulfil orders & collect products in a supermarket & Teamwork Games \\
\cite{sidji2024human} & Game & 1 human 1 AI & Codenames & win the game & codenames game & Teamwork Games \\
\cite{islam2025human} & Aviation/Military/Space & 1 human 5 AI & Simulation & teach drone how to move & drone maneovering & Aerialbased \\
\cite{bienefeld2023human} & Healthcare & 4 humans 1 AI & Simulation & diagnose and provide treatment & diagnosing patients & Workbased Activity \\
\cite{momose2025human} & Aviation/Military/Space & 1 human 1 AI & Moon landing game & safely land the space craft & landing a space craft game & Space \\
\cite{schadd2022m} & Disaster Response & 1 human 1 AI & Simulation & rescue victims & rescue victims & Search and Rescue \\
\cite{bhambri2023incorporating} & Game & 1 human 1 AI & Overcooked & deliver soup & cooking game & Teamwork Games \\
\cite{zhang2023investigating} & Game & 1 human 1 AI & Arma III & collect as many crates as possible & Arma III crate collection game & Teamwork Games \\
\cite{ye2022modeling} & Puzzles & 4 humans 1 AI & Experiment & score as high as possible & multiple choice quiz & Quiz \\
\cite{li2024modeling} & Aviation/Military/Space & 1 human 1 AI & Simulation & perform successful maintenance & spaceship maintenance & Space \\
\cite{attig2024more} & Game & 1 human 1 AI & Hanabi & win the game & Hanabi & Teamwork Games \\
\cite{zhang2024mutual} & Work & 1 human 1 AI & Overcooked & win the game & cooking game & Teamwork Games \\
\cite{smith2025navigating} & Cybersecurity/Security & 1 human 1 AI & Simulation & allow entry for authorised personnel only & facial recognition & Object Classification \\
\cite{darban2024navigating} & Education & 6-8 humans 1 AI & Experiment & complete coursework & management coursework & Workbased Activity \\
\cite{babbar2022utility} & Classification & 1 human 1 AI & Experiment & classify objects & classifying objects & Object Classification \\
\cite{wurfel2024operationalizing} & Aviation/Military/Space & 2 humans 1 AI & Simulation & safely land the plane & emergency aircraft & Aerialbased \\
\cite{lu2023readingquizmaker} & Education & 1 human 1 AI & Simulation & design a quiz & design a quiz & Workbased Activity \\
\cite{prieto2023single} & Education & 1 human 1 AI & Experiment & track thesis progression & diary for thesis progression & Workbased Activity \\
\cite{harris2023social} & Puzzles & 2-3 humans 1 AI & Simulation & solve problems & problem solving; creativity task & Workbased Activity \\
\cite{milella2023impact} & Puzzles & 3-4 humans 1 AI & Quiz & win the game & multiple choice quiz & Quiz \\
\cite{mallick2024pursuit} & Game & 1 human 5-7 AI & Netrek & win the game & destroy enemies in Netrek & Teamwork Games \\
\cite{zhao2025role} & Disaster Response & 1 human 1 AI & Simulation & rescue victims & rescue victims & Search and Rescue \\
\cite{schelble2024towards} & Aviation/Military/Space & 2 humans 1 AI & Arma III & eliminate targets & search and destroy & Aerialbased \\
\cite{jia2021towards} & Classification & 1 human 1 AI & Experiment & categorise birds correctly & classifying objects & Object Classification \\
\cite{marrone2025understanding} & Education & 2-5 humans 1 AI & Coursework & build space craft and fly it successfully & build spacecraft & Space \\
\cite{duan2024understanding} & Aviation/Military/Space & 1 human 2 AI; 2 human 1 AI & Simulation & complete surveillance & reconnaissance photography & Aerialbased \\
\cite{hauptman2024understanding} & Education & 1 human 1 AI & Packet Tracer & learn about network connections & Learn about computer networking & Workbased Activity \\
\cite{duan2025understanding} & Aviation/Military/Space & 1 human 2 AI; 2 human 1 AI & Simulation & complete surveillance & reconnaissance photography & Aerialbased \\
\cite{bansal2019updates} & Experimental Game & 1 human 1 AI & CAJA & identify defective vs non-defective boxes & classifying objects & Object Classification \\
\cite{zhang2024verbal} & Game & 1 human 2 AI; 2 human 1 AI & Rocket League & win the game & score in rocket league & Teamwork Games \\
\cite{hong2023visualizing} & Education & 1 human 1 AI & Experiment & identify cell lineages & cell lineage & Workbased Activity \\
\cite{mallick2024you} & Game & 1 human 5-7 AI & Netrek & win the game & destroy enemies in Netrek & Teamwork Games \\
\cite{mcneese2021my} & Disaster Response & 1 human 2 AI; 2 human 1 AI & NeoCITIES & deploy rescources for disaster relief & deploy resources for disaster relief & Search and Rescue \\
\cite{georganta2024would} & Work & 1 human 1 AI & Experiment & create a new fitness app & decide app functionalities & Workbased Activity \\
\cite{zhang2022you} & Classification & 1 human 1 AI & Experiment & classify objects & classifying objects & Object Classification \\
\cite{erengin2024you} & Cybersecurity/Security & 2 humans 1 AI & Simulation & correctly identify suspicious luggage & airport security Xray & Security \\ \bottomrule
\end{tabular}
}
\caption{A summary of the papers analysed, in terms of application domain, team makeup, study environment, team goal, and tasks undertaken.}
\label{tab:over}
\end{sidewaystable}

There are some strong areas of focus within Human-AI teaming experimental studies.
For example, gaming applications were a common domain studies wished to provide insight into (10 papers), alongside classification (8 papers) and aviation/military/space (7 papers).
Other common applications include puzzles (such as multiple choice quizzes; 6 papers), education (6 papers), disaster response (such as search and rescue; 5 papers), and cybersecurity/security (5 papers).
Experiment/simulation setups were a common way to study Human-AI teaming (29 papers), as well as commercial gaming environments (17 papers) such as \textit{Overcooked} (2 papers) and \textit{Rocket League} (3 papers). 
Experimental games/platforms included bespoke environments made for the purposes of the study (6 papers), and education refers to studies conducted in an educational setting (e.g. coursework; 2 papers).

The types of tasks and team goals similarly followed the trends in application domain; teamwork games (16 papers) typically involved the goal of winning the game (20 papers), and object classification tasks (12 papers) were typically associated with the goal of classifying objects (16 papers).
Tasks and goals were not fully aligned in all cases for several reasons; for example, some security tasks were associated to object classification (e.g. facial recognition), and some teamwork games were related to goals such as rescuing victims.

Finally, there was a strong preference for Human-AI teams to involve one human and one AI (34 papers).
The next largest combination was two humans with one AI (7 papers), followed by one human with two AI (6 papers) and more than three humans with one AI (6 papers).
Interestingly, only one side of the team would vary in number; if there were multiple humans, there would only be one AI, and vice versa.

Overall, the most common way Human-AI teaming has been experimentally studied is one human with one AI, in a gaming application and environment, involving teamwork games where the goal is to win the game.
Other common setups included classifying objects within simulated environments.

\subsection{Task Level Characteristics}

Within the tasks studied in experimental studies, there were a variety of roles humans and AI were assigned.
These are shown in Table \ref{tab:role}.

\begin{sidewaystable}[!]
\resizebox{\textwidth}{!}{
\begin{tabular}{@{}lll@{}}
\toprule
\textbf{Paper} & \textbf{human role} & \textbf{AI role} \\ \midrule
\cite{bansal2019case} & accept or reject recommendation; make final decision & recommends label; has access to hidden information \\
\cite{olla2024cybersecurity} & create filter rule, or accept recommendation; make final decision & recommends filter \\
\cite{amresh2023minecraft} & adhere or disregard guidance, ask for further guidance; make final decision & advise \\
\cite{schwalb2022study} & control or observe drone; search for hostages; rescue hostages; make final decision & detect hostages; mark map locations; navigate environment \\
\cite{tariq2025a2c} & review information; address detected threat; ask for help from AI; make final decision; discuss evidence with AI & log suspicious activity; discuss evidence with human; provide contextual information to human \\
\cite{josephs2023artifact} & rate confidence in deepfake; modify answer based on model prediction; make final decision & predict deepfake \\
\cite{ong2012closing} & place teammate & destroy enemies \\
\cite{zvelebilova2024collective} & solve puzzle & provide clue \\
\cite{xu2023comparing} & accept or reject recommendation; edit recommendation; make final decision & recommend box boundary \\
\cite{hemmer2025complementarity} & provide initial prediction; accept or adjust AI prediction; classify images; make final decision & provide recommendation \\
\cite{hemmer2025complementarity} & provide initial prediction; accept or adjust AI prediction; classify images; make final decision & provide recommendation \\
\cite{zhang2024crew} & provide positive or negative feedback on AI behaviour & accept feedback from human \\
\cite{munyaka2023decision} & discuss clues; provide clue to AI & guess the word from clues given by humans; provide clue to AI; discuss clues \\
\cite{mahmood2024designing} & make final decision; consider AI prediction & provide initial recommendation \\
\cite{newn2019designing} & build connections between cities; consider AI's prediction & AI infers opponent's actions \\
\cite{bansal2021does} & assign sentiment; consider AI prediction; make final decision & provide confidence score; provide recommendation \\
\cite{flathmann2024empirically} & score goals & score goals \\
\cite{flathmann2023examining} & score goals & score goals \\
\cite{jorge2024should} & collect items from aisles; deliver items to AI & provide order information \\
\cite{sidji2024human} & generate a clue; consider AI suggestion & provide clue suggestions; advise \\
\cite{islam2025human} & demonstrate movements to AI; correct AI when necessary & follow demonstration \\
\cite{bienefeld2023human} & diagnose patient; select treatment; consult AI & provide diagnostic suggestions; ventilate patient automatically \\
\cite{momose2025human} & control speed; control rotation; follow path; complete n-back test & control speed; control rotation; calculate initial path; provide confidence score; ask for help \\
\cite{schadd2022m} & assess victims; aid victims; replace robot battery; carry victim to safety & clear rubble; enter buildings; carry victim to safety \\
\cite{bhambri2023incorporating} & select strategy for agent; adapt to agents strategy; follow plan & follow strategy \\
\cite{zhang2023investigating} & collect crates & collect crates \\
\cite{ye2022modeling} & make initial decision; decide if to invoke AI; make final group decision & provide suggestion \\
\cite{li2024modeling} & accept or reject recommendation & recommend procedure \\
\cite{attig2024more} & provide hint; discard a card; play a card & provide hint; discard a card; play a card \\
\cite{zhang2024mutual} & request items; prepare food & prepare food; respond to requests \\
\cite{smith2025navigating} & compare face to record; grant or deny entry; accept or reject recommendation & provide recommendation \\
\cite{darban2024navigating} & track project progress; generate ideas; summarise sources; draft interviews; draft reports; prepare presentation & answer prompts \\
\cite{babbar2022utility} & make final decision; consider AI prediction & suggest label; provide confidence score \\
\cite{wurfel2024operationalizing} & follow known procedures; consider AI suggestion; land plane at airport & suggest alternate airports \\
\cite{lu2023readingquizmaker} & review AI suggestions; edit AI suggestions; create questions & create question suggestions \\
\cite{prieto2023single} & define problems; report progress; discuss findings; suggest new problems & analyse data; model behaviour \\
\cite{harris2023social} & solve puzzle; discuss with others & offer suggestions \\
\cite{milella2023impact} & make initial decision; consider AI decision; make final group decision & suggest answer \\
\cite{mallick2024pursuit} & protect planets; eliminate enemies & protect planets; eliminate enemies; capture planets \\
\cite{zhao2025role} & follow AI & guide human to victims \\
\cite{schelble2024towards} & ground; surveillance & aerial; move; destroy enemies; update surveillance \\
\cite{jia2021towards} & assign AI attributes & guide creation of attributes \\
\cite{marrone2025understanding} & build rocket; consider AI suggestions & provide feedback; provide progress \\
\cite{duan2024understanding} & take photos; control aircraft; provide target airspeed and altitude; create flight plan; provide waypoint information; confirm photos & create flight plan; provide waypoint information; confirm photos \\
\cite{hauptman2024understanding} & power devices; move cables & device configuration \\
\cite{duan2025understanding} & take photos; control aircraft; provide target airspeed and altitude; create flight plan; provide waypoint information; confirm photos & create flight plan; provide waypoint information; confirm photos \\
\cite{bansal2019updates} & accept or reject recommendation; make final decision & recommends label; has access to hidden information \\
\cite{zhang2024verbal} & score goals & score goals \\
\cite{hong2023visualizing} & assign 64-cell embryo; consider AI predictions & provide predictions \\
\cite{mallick2024you} & protect planets; eliminate enemies & protect planets; eliminate enemies; capture planets \\
\cite{mcneese2021my} & allocate resources; police response; fire response & hazmat response; fire response \\
\cite{georganta2024would} & designer; discuss app features & software developer; discuss app features \\
\cite{zhang2022you} & make initial decision; consider AI decision; make final decision & make recommendations \\
\cite{erengin2024you} & identify luggage & discuss capabilities; identify luggage \\ \bottomrule
\end{tabular}
}
\caption{An overview of the human and AI roles found within experimental studies of human-AI teams.}
\label{tab:role}
\end{sidewaystable}

\begin{figure}[h]
    \centering
    \includegraphics[width=\linewidth]{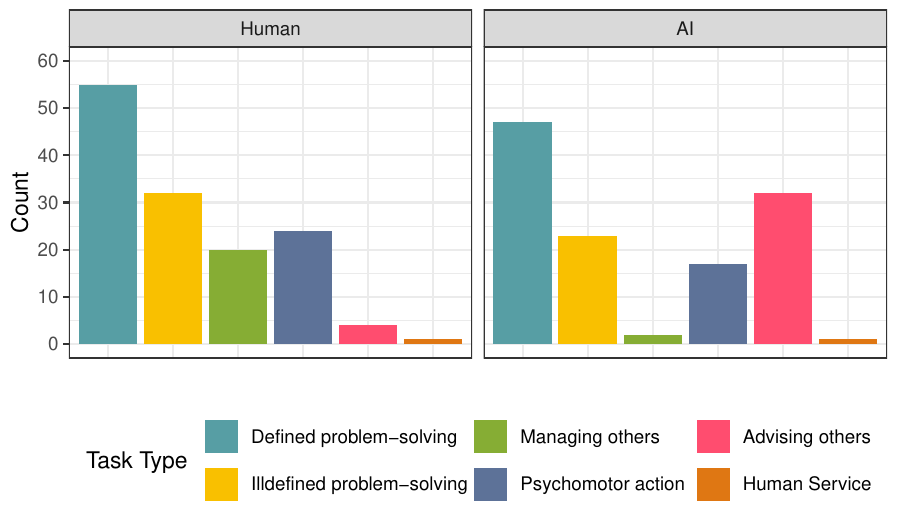}
    \caption{A bar chart showing the different task level characteristics for each paper, separated by human and AI teammates.}
    \label{fig:tasktype}
\end{figure}

There is a large variety in roles assigned to both humans and AI, though a common role for human teammates was to accept/reject an AI recommendation to make a final decision (18 studies), whilst a common role for AI was to provide a recommendation/advise a human teammate (15 studies).
The types of tasks are classified in the task level characteristic taxonomy from \cite{wildman2012task} in Table \ref{tab:task}, and summarised in Figure \ref{fig:tasktype}.

\begin{sidewaystable}[!]
\resizebox{\textwidth}{!}{
\begin{tabular}{@{}ccccccclcccccc@{}}
\cmidrule(r){1-7} \cmidrule(l){9-14}
 & \multicolumn{6}{c}{Human Teammate(s)} &  & \multicolumn{6}{c}{AI Teammates(s)} \\
\textbf{Paper} & \textbf{\begin{tabular}[c]{@{}c@{}}Defined problem\\ solving\end{tabular}} & \textbf{\begin{tabular}[c]{@{}c@{}}Ill-defined problem\\ solving\end{tabular}} & \textbf{\begin{tabular}[c]{@{}c@{}}Psychomotor\\ action\end{tabular}} & \textbf{\begin{tabular}[c]{@{}c@{}}Managing\\ others\end{tabular}} & \textbf{\begin{tabular}[c]{@{}c@{}}Advising\\ others\end{tabular}} & \textbf{\begin{tabular}[c]{@{}c@{}}Human\\ Service\end{tabular}} &  & \textbf{\begin{tabular}[c]{@{}c@{}}Defined problem\\ solving\end{tabular}} & \textbf{\begin{tabular}[c]{@{}c@{}}Ill-defined problem\\ solving\end{tabular}} & \textbf{\begin{tabular}[c]{@{}c@{}}Psychomotor\\ action\end{tabular}} & \textbf{\begin{tabular}[c]{@{}c@{}}Managing\\ others\end{tabular}} & \textbf{\begin{tabular}[c]{@{}c@{}}Advising\\ others\end{tabular}} & \textbf{\begin{tabular}[c]{@{}c@{}}Human\\ service\end{tabular}} \\ \cmidrule(r){1-7} \cmidrule(l){9-14} 
\cite{bansal2019case} & \checkmark &  &  &  &  &  &  & \checkmark &  &  &  & \checkmark &  \\
\cite{olla2024cybersecurity} & \checkmark &  &  & \checkmark &  &  &  & \checkmark &  &  &  & \checkmark &  \\
\cite{amresh2023minecraft} & \checkmark & \checkmark & \checkmark &  &  &  &  & \checkmark & \checkmark &  &  & \checkmark &  \\
\cite{schwalb2022study} & \checkmark & \checkmark & \checkmark & \checkmark &  &  &  & \checkmark & \checkmark &  &  & \checkmark &  \\
\cite{tariq2025a2c} & \checkmark &  &  & \checkmark &  &  &  & \checkmark &  &  &  & \checkmark &  \\
\cite{josephs2023artifact} & \checkmark &  &  &  &  &  &  & \checkmark &  &  &  & \checkmark &  \\
\cite{ong2012closing} & \checkmark & \checkmark & \checkmark & \checkmark &  &  &  & \checkmark &  & \checkmark &  &  &  \\
\cite{zvelebilova2024collective} & \checkmark & \checkmark &  &  &  &  &  &  &  &  &  & \checkmark &  \\
\cite{xu2023comparing} & \checkmark &  &  & \checkmark &  &  &  & \checkmark &  &  &  & \checkmark &  \\
\cite{hemmer2025complementarity} & \checkmark &  &  & \checkmark &  &  &  & \checkmark &  &  &  & \checkmark &  \\
\cite{hemmer2025complementarity} & \checkmark &  &  & \checkmark &  &  &  & \checkmark &  &  &  & \checkmark &  \\
\cite{zhang2024crew} & \checkmark &  &  & \checkmark &  &  &  & \checkmark &  &  &  &  &  \\
\cite{munyaka2023decision} & \checkmark & \checkmark &  &  &  &  &  & \checkmark & \checkmark &  &  & \checkmark &  \\
\cite{mahmood2024designing} & \checkmark &  &  &  &  &  &  & \checkmark &  &  &  & \checkmark &  \\
\cite{newn2019designing} & \checkmark & \checkmark & \checkmark &  &  &  &  &  & \checkmark &  &  & \checkmark &  \\
\cite{bansal2021does} & \checkmark &  &  &  &  &  &  & \checkmark &  &  &  & \checkmark &  \\
\cite{flathmann2024empirically} & \checkmark & \checkmark & \checkmark &  &  &  &  & \checkmark & \checkmark & \checkmark &  &  &  \\
\cite{flathmann2023examining} & \checkmark & \checkmark & \checkmark &  &  &  &  & \checkmark & \checkmark & \checkmark &  &  &  \\
\cite{jorge2024should} & \checkmark & \checkmark & \checkmark &  &  &  &  &  &  &  & \checkmark &  &  \\
\cite{sidji2024human} & \checkmark & \checkmark &  &  &  &  &  &  & \checkmark &  &  & \checkmark &  \\
\cite{islam2025human} & \checkmark & \checkmark & \checkmark & \checkmark &  &  &  & \checkmark & \checkmark & \checkmark &  &  &  \\
\cite{bienefeld2023human} & \checkmark & \checkmark & \checkmark & \checkmark & \checkmark &  &  & \checkmark & \checkmark & \checkmark &  & \checkmark &  \\
\cite{momose2025human} & \checkmark & \checkmark & \checkmark & \checkmark &  &  &  & \checkmark & \checkmark & \checkmark &  & \checkmark &  \\
\cite{schadd2022m} & \checkmark & \checkmark & \checkmark & \checkmark &  & \checkmark &  & \checkmark & \checkmark & \checkmark &  &  &  \\
\cite{bhambri2023incorporating} & \checkmark & \checkmark & \checkmark & \checkmark &  &  &  & \checkmark & \checkmark & \checkmark &  &  &  \\
\cite{zhang2023investigating} & \checkmark & \checkmark & \checkmark &  &  &  &  & \checkmark & \checkmark & \checkmark &  &  &  \\
\cite{ye2022modeling} & \checkmark &  &  &  &  &  &  & \checkmark &  &  &  & \checkmark &  \\
\cite{li2024modeling} & \checkmark &  &  &  &  &  &  & \checkmark &  &  &  & \checkmark &  \\
\cite{attig2024more} & \checkmark &  &  &  &  &  &  & \checkmark &  &  &  &  &  \\
\cite{zhang2024mutual} & \checkmark &  &  & \checkmark &  &  &  & \checkmark &  &  &  &  &  \\
\cite{smith2025navigating} & \checkmark &  &  &  &  &  &  & \checkmark &  &  &  & \checkmark &  \\
\cite{darban2024navigating} & \checkmark & \checkmark &  & \checkmark & \checkmark &  &  &  &  &  &  & \checkmark & \checkmark \\
\cite{babbar2022utility} & \checkmark &  &  &  &  &  &  & \checkmark &  &  &  & \checkmark &  \\
\cite{wurfel2024operationalizing} & \checkmark & \checkmark & \checkmark & \checkmark & \checkmark &  &  & \checkmark &  & \checkmark &  & \checkmark &  \\
\cite{lu2023readingquizmaker} & \checkmark & \checkmark &  & \checkmark &  &  &  & \checkmark & \checkmark &  &  & \checkmark &  \\
\cite{prieto2023single} & \checkmark & \checkmark &  & \checkmark & \checkmark &  &  &  &  &  &  & \checkmark &  \\
\cite{harris2023social} & \checkmark & \checkmark &  &  &  &  &  & \checkmark & \checkmark &  &  & \checkmark &  \\
\cite{milella2023impact} & \checkmark &  &  &  &  &  &  & \checkmark &  &  &  & \checkmark &  \\
\cite{mallick2024pursuit} & \checkmark & \checkmark & \checkmark &  &  &  &  & \checkmark & \checkmark & \checkmark &  &  &  \\
\cite{zhao2025role} &  &  & \checkmark &  &  &  &  & \checkmark &  &  & \checkmark &  &  \\
\cite{schelble2024towards} & \checkmark & \checkmark & \checkmark & \checkmark &  &  &  & \checkmark & \checkmark & \checkmark &  &  &  \\
\cite{jia2021towards} & \checkmark &  &  & \checkmark &  &  &  & \checkmark &  &  &  & \checkmark &  \\
\cite{marrone2025understanding} & \checkmark &  &  &  &  &  &  & \checkmark &  &  &  & \checkmark &  \\
\cite{duan2024understanding} & \checkmark & \checkmark & \checkmark &  &  &  &  & \checkmark & \checkmark & \checkmark &  &  &  \\
\cite{hauptman2024understanding} & \checkmark &  & \checkmark &  &  &  &  & \checkmark &  & \checkmark &  &  &  \\
\cite{duan2025understanding} & \checkmark & \checkmark & \checkmark &  &  &  &  & \checkmark & \checkmark & \checkmark &  &  &  \\
\cite{bansal2019updates} & \checkmark &  &  &  &  &  &  & \checkmark &  &  &  & \checkmark &  \\
\cite{zhang2024verbal} & \checkmark & \checkmark & \checkmark &  &  &  &  & \checkmark & \checkmark & \checkmark &  &  &  \\
\cite{hong2023visualizing} & \checkmark &  &  &  &  &  &  & \checkmark &  &  &  & \checkmark &  \\
\cite{mallick2024you} & \checkmark & \checkmark & \checkmark &  &  &  &  & \checkmark & \checkmark & \checkmark &  &  &  \\
\cite{mcneese2021my} & \checkmark & \checkmark &  &  &  &  &  & \checkmark & \checkmark &  &  &  &  \\
\cite{georganta2024would} &  & \checkmark &  &  &  &  &  &  & \checkmark &  &  &  &  \\
\cite{zhang2022you} & \checkmark &  &  &  &  &  &  & \checkmark &  &  &  & \checkmark &  \\
\cite{erengin2024you} & \checkmark &  &  &  &  &  &  & \checkmark &  &  &  &  &  \\ \cmidrule(r){1-7} \cmidrule(l){9-14} 
\end{tabular}
}
\caption{A table showing the task level characteristics of both humans and AI in experimental papers. Note that papers 146, 151, 166 and 176 contain no AI.}
\label{tab:task}
\end{sidewaystable}

The majority of studies involve defined problem-solving for both humans (52 studies) and AI (47 studies), and a sizeable portion involve ill-defined problem-solving (human roles 29 studies, AI roles 23 studies) and psychomotor action (human roles 22 studies, AI roles 17 studies).
Human service was very rare for both humans and AI (one instance each), and there were no instances of negotiation as there were no team dynamics in which humans and AI intentionally held opposing views.
The main difference between roles was in managing vs advising others, with the majority of human roles involving managing (20 studies) rather than advising (4 studies), and the opposite trend for AI (2 managing vs 32 advising studies).
Overall, the majority of studies on Human-AI teaming involve humans and AI engaged in defined and ill-defined problem solving, psychomotor action, where the human takes a managing role whilst the AI takes an advisory role.

\subsection{Team Level Characteristics}

\begin{figure}[h]
    \centering
    \includegraphics[width=.9\linewidth]{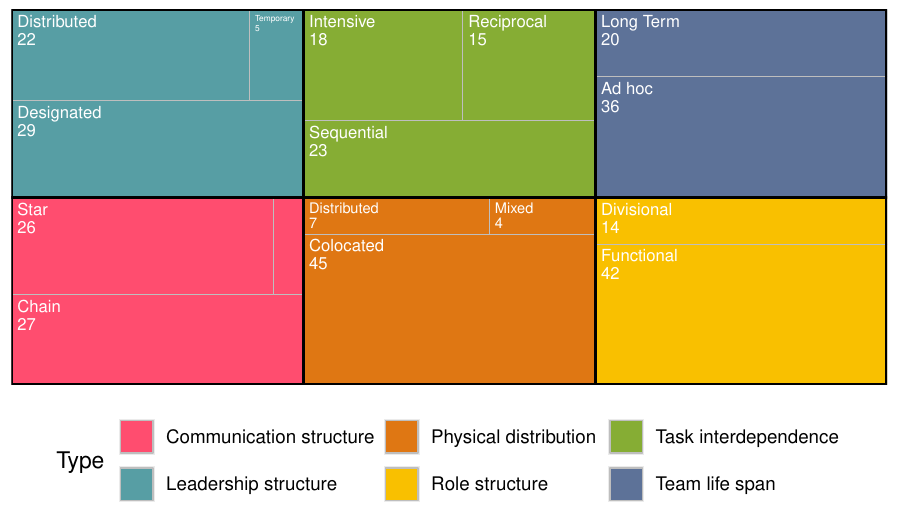}
    \caption{A tree diagram illustrating how each of the six team level characteristics break down.}
    \label{fig:teamtype}
\end{figure}

Experimental studies varied in terms of the types of team level characteristics, shown in Table \ref{tab:team} and summarised in Figure \ref{fig:teamtype}.
\begin{table}[h]
\resizebox{\textwidth}{!}{
\begin{tabular}{@{}ccccccc@{}}
\toprule
\textbf{Paper} & \textbf{Task interdependence} & \textbf{Role structure} & \textbf{Leadership structure} & \textbf{Communication structure} & \textbf{Physical distribution} & \textbf{Team life span} \\ \midrule
\cite{bansal2019case} & Sequential & Functional & Designated & Chain & Colocated & Long Term \\
\cite{olla2024cybersecurity} & Sequential & Functional & Designated & Chain & Colocated & Long Term \\
\cite{amresh2023minecraft} & Intensive & Functional & Distributed & Star & Mixed & Ad hoc \\
\cite{schwalb2022study} & Reciprocal & Functional & Designated & Chain & Colocated & Ad hoc \\
\cite{tariq2025a2c} & Reciprocal & Functional & Designated & Chain & Colocated & Long Term \\
\cite{josephs2023artifact} & Sequential & Functional & Designated & Chain & Colocated & Long Term \\
\cite{ong2012closing} & Sequential & Functional & Designated & Chain & Colocated & Ad hoc \\
\cite{zvelebilova2024collective} & Intensive & Functional & Distributed & Star & Mixed & Ad hoc \\
\cite{xu2023comparing} & Sequential & Functional & Designated & Chain & Colocated & Long Term \\
\cite{hemmer2025complementarity} & Sequential & Functional & Designated & Chain & Colocated & Long Term \\
\cite{hemmer2025complementarity} & Sequential & Functional & Designated & Chain & Colocated & Long Term \\
\cite{zhang2024crew} & Sequential & Functional & Designated & Chain & Colocated & Ad hoc \\
\cite{munyaka2023decision} & Intensive & Functional & Distributed & Hub-and-Wheel & Colocated & Ad hoc \\
\cite{mahmood2024designing} & Sequential & Functional & Designated & Chain & Colocated & Long Term \\
\cite{newn2019designing} & Reciprocal & Functional & Designated & Chain & Colocated & Ad hoc \\
\cite{bansal2021does} & Sequential & Functional & Designated & Chain & Colocated & Long Term \\
\cite{flathmann2024empirically} & Reciprocal & Divisional & Distributed & Star & Colocated & Ad hoc \\
\cite{flathmann2023examining} & Reciprocal & Divisional & Distributed & Star & Colocated & Ad hoc \\
\cite{jorge2024should} & Sequential & Functional & Designated & Hub-and-Wheel & Colocated & Long Term \\
\cite{sidji2024human} & Sequential & Divisional & Designated & Chain & Colocated & Ad hoc \\
\cite{islam2025human} & Intensive & Functional & Designated & Hub-and-Wheel & Colocated & Ad hoc \\
\cite{bienefeld2023human} & Intensive & Functional & Distributed & Star & Colocated & Ad hoc \\
\cite{momose2025human} & Reciprocal & Divisional & Distributed & Star & Colocated & Ad hoc \\
\cite{schadd2022m} & Reciprocal & Functional & Distributed & Star & Colocated & Ad hoc \\
\cite{bhambri2023incorporating} & Reciprocal & Functional & Designated & Chain & Colocated & Ad hoc \\
\cite{zhang2023investigating} & Reciprocal & Divisional & Distributed & Star & Colocated & Ad hoc \\
\cite{ye2022modeling} & Sequential & Functional & Distributed & Star & Colocated & Ad hoc \\
\cite{li2024modeling} & Reciprocal & Divisional & Designated & Chain & Colocated & Ad hoc \\
\cite{attig2024more} & Sequential & Divisional & Temporary & Star & Colocated & Ad hoc \\
\cite{zhang2024mutual} & Reciprocal & Functional & Designated & Chain & Colocated & Ad hoc \\
\cite{smith2025navigating} & Sequential & Functional & Designated & Chain & Colocated & Long Term \\
\cite{darban2024navigating} & Intensive & Divisional & Distributed & Star & Mixed & Long Term \\
\cite{babbar2022utility} & Sequential & Functional & Designated & Chain & Colocated & Ad hoc \\
\cite{wurfel2024operationalizing} & Intensive & Functional & Temporary & Star & Colocated & Ad hoc \\
\cite{lu2023readingquizmaker} & Reciprocal & Divisional & Designated & Chain & Colocated & Long Term \\
\cite{prieto2023single} & Sequential & Functional & Designated & Chain & Colocated & Long Term \\
\cite{harris2023social} & Intensive & Divisional & Temporary & Star & Distributed & Ad hoc \\
\cite{milella2023impact} & Intensive & Divisional & Distributed & Star & Mixed & Ad hoc \\
\cite{mallick2024pursuit} & Intensive & Functional & Distributed & Star & Colocated & Ad hoc \\
\cite{zhao2025role} & Sequential & Functional & Designated & Chain & Distributed & Ad hoc \\
\cite{schelble2024towards} & Intensive & Functional & Distributed & Star & Distributed & Ad hoc \\
\cite{jia2021towards} & Sequential & Functional & Designated & Chain & Colocated & Long Term \\
\cite{marrone2025understanding} & Intensive & Functional & Distributed & Star & Colocated & Ad hoc \\
\cite{duan2024understanding} & Intensive & Functional & Distributed & Star & Distributed & Ad hoc \\
\cite{hauptman2024understanding} & Reciprocal & Functional & Designated & Chain & Colocated & Ad hoc \\
\cite{duan2025understanding} & Intensive & Functional & Distributed & Star & Distributed & Ad hoc \\
\cite{bansal2019updates} & Sequential & Functional & Designated & Chain & Colocated & Long Term \\
\cite{zhang2024verbal} & Intensive & Divisional & Distributed & Star & Colocated & Ad hoc \\
\cite{hong2023visualizing} & Sequential & Functional & Designated & Chain & Colocated & Long Term \\
\cite{mallick2024you} & Intensive & Functional & Distributed & Star & Colocated & Ad hoc \\
\cite{mcneese2021my} & Intensive & Functional & Distributed & Star & Distributed & Ad hoc \\
\cite{georganta2024would} & Reciprocal & Functional & Distributed & Star & Distributed & Long Term \\
\cite{zhang2022you} & Sequential & Functional & Designated & Chain & Colocated & Long Term \\
\cite{erengin2024you} & Intensive & Functional & Distributed & Star & Colocated & Long Term \\ \bottomrule
\end{tabular}
}
\caption{A table showing the team level characteristics of the papers studied. Note that 166 and 176 are empty as these experiments contained no AI and only one person.}
\label{tab:team}
\end{table}
The majority of discrete categories were present in the experimental studies, except for pooled interdependence, and external manager leadership structure.
This is likely because all Human-AI Teaming scenarios involved a leader who was present inside the team, and there was always a level of interaction between team members.

As can be seen, some categories were more common than others, such as a preference for functional role structures (i.e. the team members do distinctly different roles that are not interchangeable).
Given that Human-AI Teaming is commonly associated with the idea that humans and AI have complementary strengths and weaknesses (e.g. \cite{mcgrath2025collaborative, berretta2023defining}), this is understandable.
However, some categories within characteristics were more varied.
For example, task interdependence was fairly evenly split between sequential, reciprocal, and intensive.
Consequently, there were a varied number of combinations of team characteristics present in the literature.
The most common combinations are explored in the following section.

\subsection{The Sub-Types of Human-AI Teams}

The five largest clusters of Human-AI Teams are described here, accounting for 41 of 53 papers (77\%).
They are presented in decreasing size order.

\subsubsection{Team Type 1: AI Assistant}

\begin{figure}[h]
    \centering
    \includegraphics[width=\linewidth]{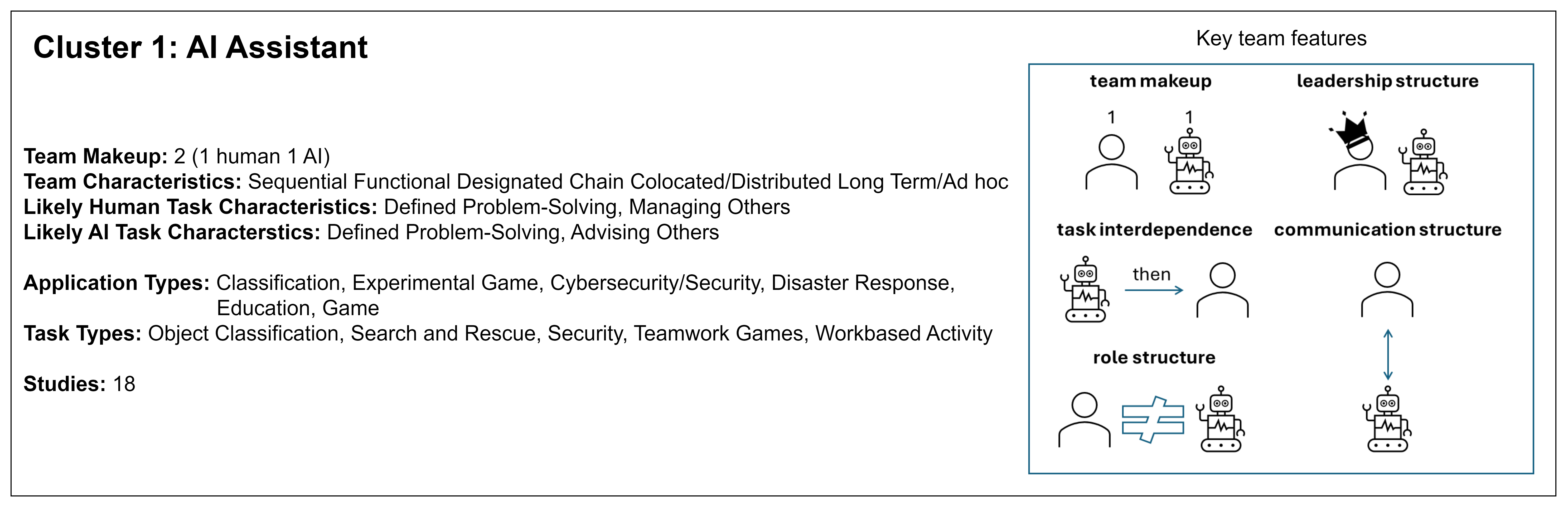}
    \caption{A summary of the AI Assistant sub-type of Human-AI Teaming.}
    \label{fig:Cluster1}
\end{figure}

The largest cluster consists of 18 studies (34\%), and are teams where a singular AI assists a singular human with a team task.
The characteristics of this cluster are summarised in Figure \ref{fig:Cluster1}.

Teams of this kind have a \textbf{sequential} task interdependence, with \textbf{functional} roles, \textbf{designated} leadership, and \textbf{chain} communication, where humans and AI are primarily \textbf{colocated} (17/18) and primarily have a \textbf{long term} team lifespan (14/18).
In other words, a human and AI take a turn in completing their part of the task, where their roles are not interchangeable, as the AI assistant has access to more information via model training and the human is the only one allowed to make a final decision.
Further, one team member remains in charge throughout, and they communicate within a hierarchical structure.

For example, in Paper \cite{mahmood2024designing}, the team is required to classify a series of images including easily recognisable objects (e.g. church, golf ball) and challenging dog breeds (e.g. old English sheepdog, Australian terrier).
To do so, an AI model recommends a classification decision, then the human also considers a recommendation before making the final team decision.
Consequently, the team members work in sequence (AI prediction followed by human decision) where it is not possible for the human and AI to swap roles.
Further, the human is designated in charge of the team decision, meaning the AI communicates its decision `up' the chain to its leader.
The team members are somewhat colocated, in that they are both working on a digital task, and the team lifespan could be considered long term if the task of classification was part of the team's `job.'

The types of tasks humans are likely to engage in within this cluster include defined problem-solving (17/18), and managing others (typically their AI teammate, 8/18), whereas AI team members are likely to engage in defined problem-solving (17/18) and advising others (i.e. their human teammate, 15/18).
For example, in Paper \cite{hemmer2025complementarity}, the team is asked to appraise a series of real estate descriptions, whereby the human first estimates the house price, followed by seeing the AI's prediction, and subsequently `adjusting' the AI's prediction to create the final prediction.
Consequently, the human and AI are both involved in defined problem-solving, however the AI is advising the human about the prediction, and the human is managing the AI by altering its output.
This is slightly in contrast to the example explained from Paper \cite{mahmood2024designing}, whereby there is no managing of the AI by the human, as they make their own judgement after seeing the AI's prediction.

Common application domains where this type of team are studied include classification tasks (8/18), experimental games (3/18), and cybersecurity/security (3/18).
Common tasks similarly involve object classification (11/18), security (2/18), teamwork games (2/18), and workbased activities (2/18).

\subsubsection{Team Type 2: Ad hoc dependency teams}

\begin{figure}[h]
    \centering
    \includegraphics[width=\linewidth]{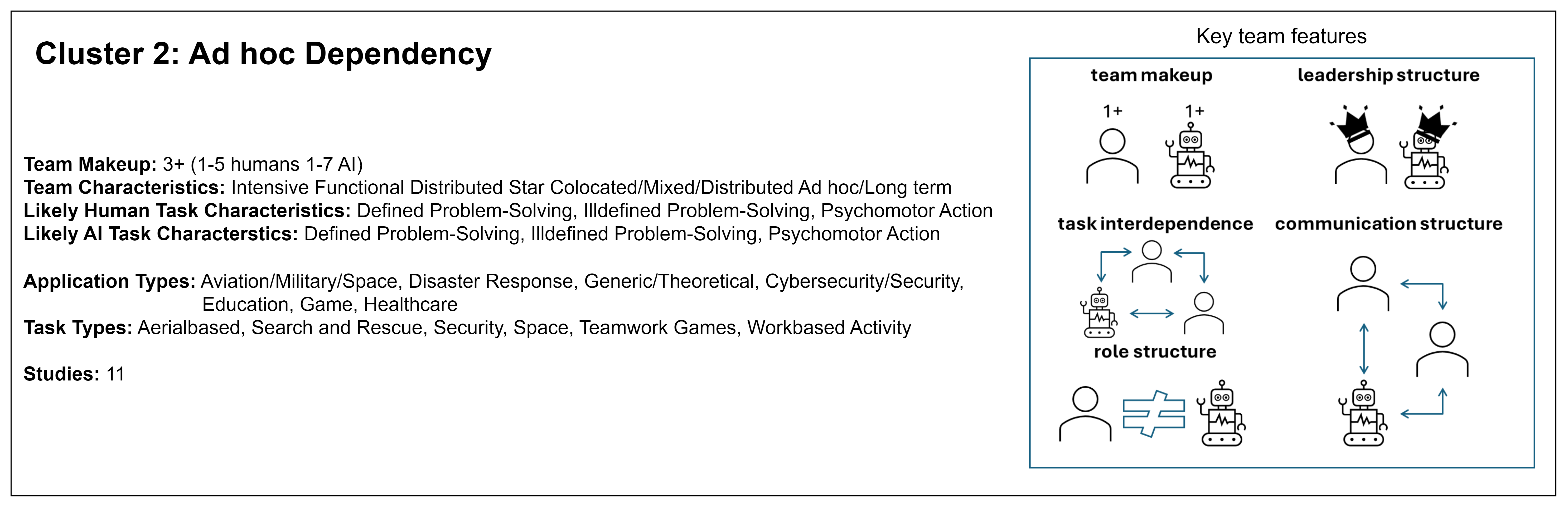}
    \caption{A summary of the Ad hoc dependency team sub-type of Human-AI Teaming.}
    \label{fig:Cluster2}
\end{figure}

The second largest cluster consists of 11 studies (20\%), and are teams where there are at least three members, with at least one team member being an AI, performing different roles together.
The characteristics of this cluster are summarised in Figure \ref{fig:Cluster2}.

Teams of this kind have an \textbf{intensive} task interdependence, with \textbf{functional} roles, \textbf{distributed} leadership, and \textbf{star} communication, where humans and AI are primarily \textbf{colocated} (5/11) or \textbf{mixed distribution} (4/11) and primarily have an \textbf{ad hoc} team lifespan (10/11).
In other words, a mix of at least three humans and AI work as a unit, where their roles are not interchangeable, leadership is shared simultaneously between team members, and they communicate freely without the need for a central point of contact.
They typically come together to perform a specific task before disbanding, but may or may not be physically located close to one another during.
Overall, the most common combination was three team members (5/11 studies), with comparable preference for multiple humans (6/11) versus multiple AI (5/11).

For example, Paper \cite{duan2024understanding} explores an aviation/military type of environment involving a team of two humans and one AI: a navigator (AI), pilot (human), and photographer (human).
Together, they performed reconnaissance photography of targets during missions.
Consequently, the team members work alongside each other intensely, where it is not possible to swap roles.
Further, leadership is distributed in that all three must lead their own portion of the task, leading to a star communication where all members communicate freely with one another.
The members would come together to perform the reconnaissance, before disbanding once the task is complete.

The types of tasks both humans and AI are likely to engage in within this cluster include defined problem-solving (11 humans vs 10 AI), ill-defined problem-solving (9 humans vs 8 AI), and psychomotor action (7 humans vs 6 AI).
For example, in Paper \cite{mallick2024pursuit} a human works alongside 5-7 AI in the game \textit{Netrek} to destroy enemy ships (human and AI role) and conquer planets (AI role only).
Consequently, both humans and AI engage in defined problem-solving to destroy targets, as well as ill-defined problem-solving in order to strategise during engagements with enemies.
Both also involve psychomotor action, in that humans and AI control ships as part of the activity.

Common application domains for this team type include aviation/military/space (3/11 papers), puzzles (2/11), and disaster response (2/11).
Common tasks similarly include aerialbased (3/11), teamwork games (3/11), and search and rescue (2/11).

\subsubsection{Team Type 3: Ad hoc forced dependency teams}

\begin{figure}[h]
    \centering
    \includegraphics[width=\linewidth]{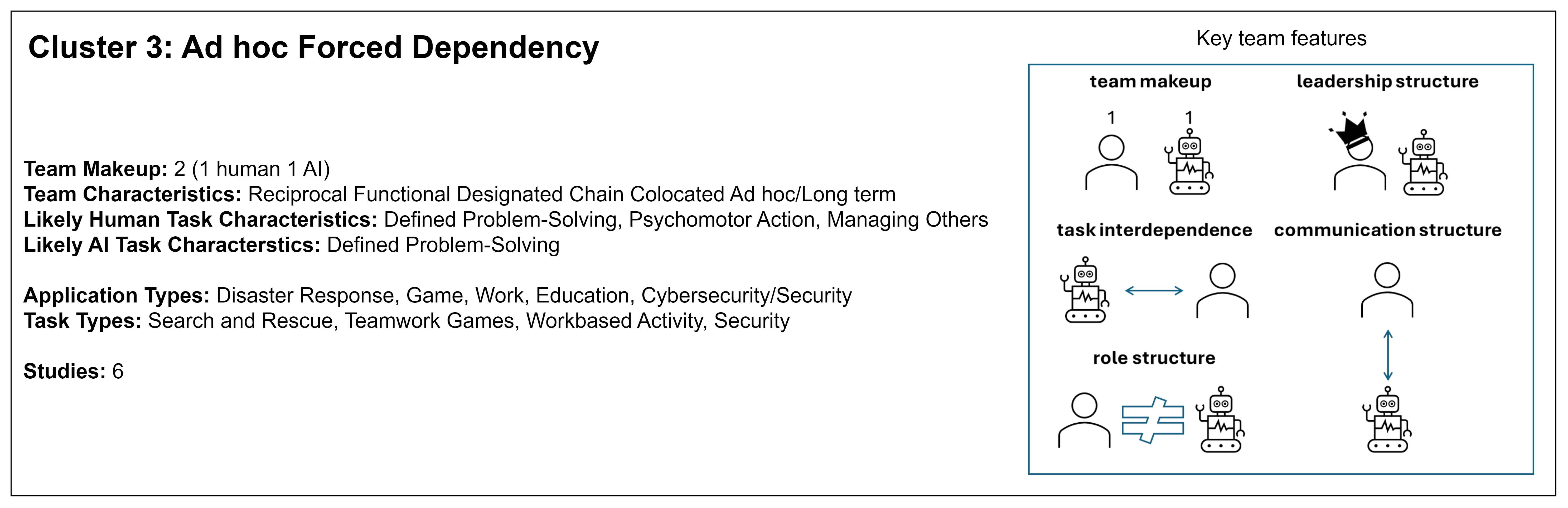}
    \caption{A summary of the Ad hoc forced dependency team sub-type of Human-AI Teaming.}
    \label{fig:Cluster3}
\end{figure}

The third largest cluster consists of six studies (11\%), and are teams with a singular human and singular AI, similar to Cluster 1.
The characteristics of this cluster are summarised in Figure \ref{fig:Cluster3}.

Teams of this kind have a \textbf{reciprocal} task interdependence, with \textbf{functional} roles, \textbf{designated} leadership, and \textbf{chain} communication, where humans and AI are \textbf{colocated} and primarily have an \textbf{ad hoc} team lifespan (5/6).
In other words, a human and AI engage in a back and forth interaction, where their roles are not interchangeable, one team member remains in charge throughout, and they communicate within a hierarchical structure.

This is therefore different than Cluster 1 (AI assistant), as teams here involve multiple stages of interacting with one another, and teams are typically formed to perform a specific task before disbanding.
They are consequently `forced' to interact with one another to achieve the team goal, typically because not all actions can be completed by one person, or one side of the team has access to information the other does not.
This is in contrast to Cluster 1, where both teammates are technically capable of completing the task on their own, but are theorised to perform better in combination.
The team type is also similar to Cluster 2 (ad hoc dependency teams), however only contains two teammates, thereby changing the interdependence, leadership, and communication style.

For example, in Paper \cite{bhambri2023incorporating}, the team is required to work together in order to successfully deliver food orders in the game \textit{Overcooked}.
Each team member has access to specific elements of the cooking process (e.g. chopping, frying), meaning they have different roles to fill.
The human is the designated leader, as the AI is instructed on what elements of the cooking is required, consequently creating a chain style of communication.
The human and AI exist within the same simulated environment, and they come together to complete the level before disbanding.

The types of tasks humans are likely to engage in within this cluster include defined problem-solving (6/6), psychomotor action (4/6), and managing others (i.e. their AI teammate, 4/6), whereas AI are likely to engage in defined problem-solving (5/6).
For example, Paper \cite{schwalb2022study} explores a combat search and rescue simulation where an AI controls a drone that surveys the area looking for hostages, which the human teammate can control whether the system is fully autonomous or not.
Consequently, both the human and AI are involved in defined problem-solving (finding hostages) and psychomotor action (controlling the drone), but the human is also involved in managing others (i.e. the AI drone) and ill-defined problem-solving (when/how to alter the drone's navigation) , whilst the AI advises the human.
This is therefore different from Cluster 1 again, as there is more likely to be psychomotor action involved.

Application domains were varied for this team type, including games (2/6), as well as disaster response, work, education, and cybersecurity/security (1 each).
Common tasks primarily involved teamwork games (3/6).

\subsubsection{Team Type 4: Paired Equanimity}

\begin{figure}[h]
    \centering
    \includegraphics[width=\linewidth]{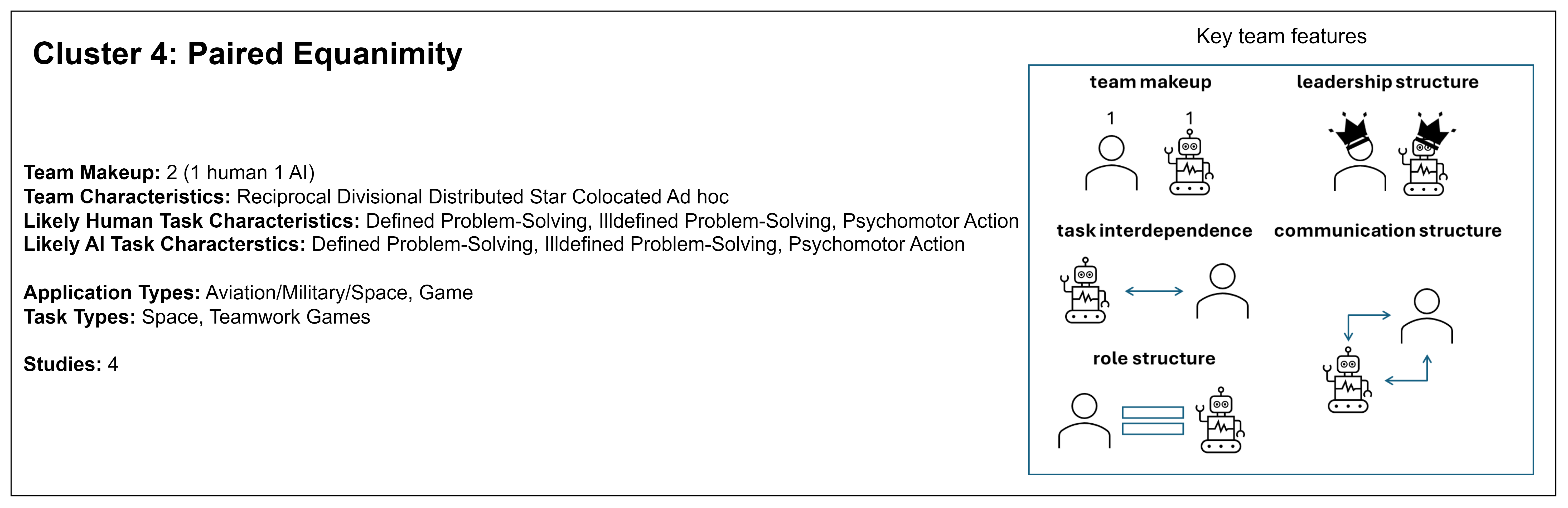}
    \caption{A summary of the paired equanimity team sub-type of Human-AI Teaming.}
    \label{fig:Cluster4}
\end{figure}

The fourth largest cluster consists of four studies (7\%), and are teams with two teammates that perform the same role together.
The characteristics of this cluster are summarised in Figure \ref{fig:Cluster4}.

Teams of this kind have a \textbf{reciprocal} task interdependence, with \textbf{divisional} roles, \textbf{distributed} leadership, and \textbf{star} communication, where humans and AI are \textbf{colocated} and have an \textbf{ad hoc} team lifespan.
In other words, a human and AI engage in a back and forth interaction, where they both perform the same role, leadership is shared simultaneously, and they communicate freely.
The teammates come together to perform a specific task before disbanding, are are colocated during the task.
This is therefore different than Cluster 3 (Ad hoc forced dependency) because here the team members share leadership and roles, which in turn changes the communication style to a star rather than a chain.

For example, in Paper \cite{flathmann2024empirically} a human and AI work together to score goals in the game \textit{Rocket League}.
Consequently, the team members work reciprocally (i.e. passing the ball back and forth) whilst performing the same role of controlling a team vehicle.
Further, leadership is distributed as both teammates can take control of the ball, leading to a star communication style.
The members come together to perform the game before disbanding, whilst both being colocated during the match.

The types of tasks humans and AI engage with overlap, as they are both likely to engage in defined problem-solving (4/4), ill-defined problem-solving (4/4), and psychomotor action (4/4).
For example, in Paper \cite{zhang2023investigating}, a human and AI work together in the combat simulation game \textit{Arma III} to collect as many boxes, in numerical order, as possible, in eight minutes.
Consequently, both human and AI engage in the same defined and ill-defined problem-solving (i.e. locating boxes whilst coordinating actions) as well as psychomotor actions, by navigating the environment.

Common application domains where this type of team are studied games (3/4) and aviation/military/space (1/4).
Similarly, the common tasks include teamwork games (3/4) and space settings (1/4).

\subsubsection{Team Type 5: Group Equanimity}

\begin{figure}[h]
    \centering
    \includegraphics[width=\linewidth]{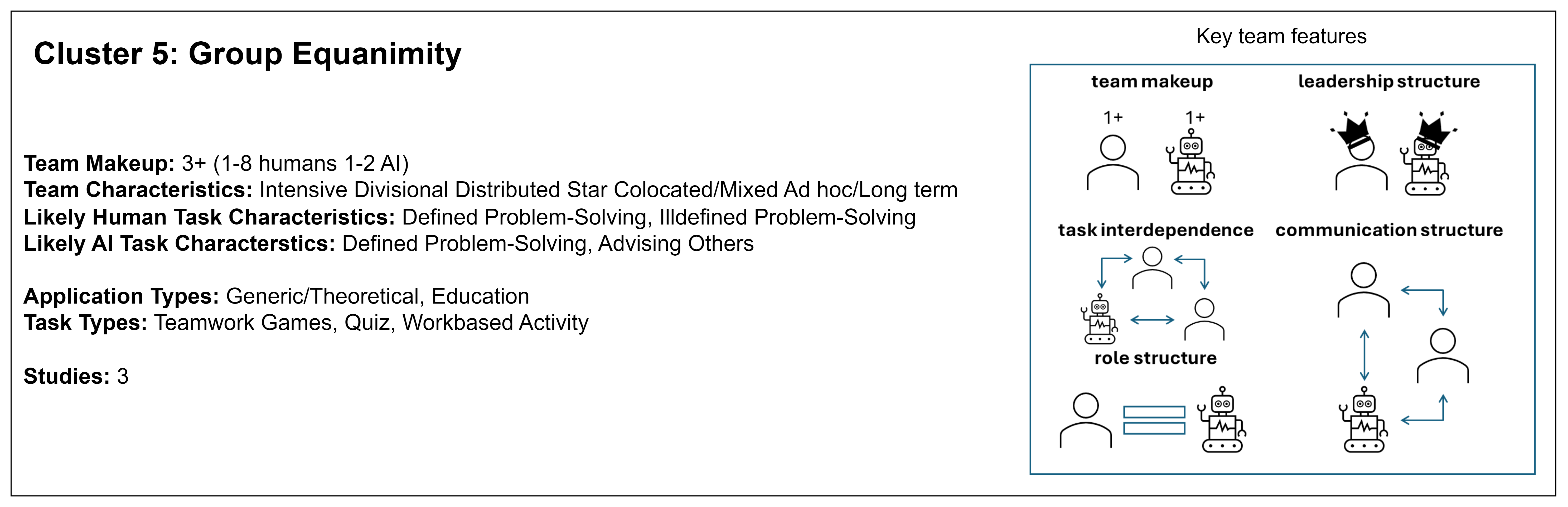}
    \caption{A summary of the Group Equanimity team sub-type of Human-AI Teaming.}
    \label{fig:Cluster5}
\end{figure}

The fifth largest cluster consists of three studies (6\%), and are teams where there are at least three members, with at least one team member being an AI, perform the same role together.
The characteristics of this cluster are summarised in Figure \ref{fig:Cluster5}.

Teams of this kind have an \textbf{intensive} task interdependence, with \textbf{divisional} roles, \textbf{distributed} leadership, and \textbf{star} communication, where humans and AI are likely \textbf{mixed distribution} (2/3 studies) and have an \textbf{ad hoc} team lifespan (2/3).
In other words, a mix of at least three humans and AI work as a unit, where they all perform the same role, leadership is shared simultaneously, and they communicate freely.
The teammates come together to perform a specific task before disbanding, are are a mixture of physically colocated and distributed during the task.
Overall, the typical team makeup was between 1-8 humans with 1-2 AI, with a slight preference for multiple humans (2/3) rather than multiple AI (1/3).

Consequently, this team type is similar to Cluster 2 (ad hoc dependency teams), however the roles being performed are the same/interchangeable.
It is also similar to Cluster 4 (paired equanimity), except there are at least three team members, which subsequently changes the task interdependence.

For example, in Paper \cite{milella2023impact}, 3-4 humans work with an AI to answer a series of quiz questions.
Each member first individually guesses the answer, before they each show their answer to the team to reach a consensus.
Consequently, the team members work alongside each other intensely, where the task of deciding the answer is jointly undertaken.
Further, leadership is distributed across team members to reach a consensus, leading to a star communication.
The members come together to answer the quiz, before disbanding.

The types of tasks humans are likely to engage with include defined problem-solving (3/3) and ill-defined problem-solving (2/3), and AI are likely to engage in defined problem-solving (2/3) and advising others (2/3).
For example, in Paper \cite{zhang2024verbal} a game of \textit{Rocket League} is played between three team members (2 humans, 1 AI), who work together to score goals.
This experimental setup is therefore different than Cluster 4 (paired equanimity), as there are three teammates rather than two.
For this experiment, both humans and AI engage with defined and ill-defined problem-solving (i.e. coordinating the scoring of goals), as well as psychomotor action by moving vehicles to interact with the ball.

Common application domains include puzzles (2/3), and education (1/3).
Common task types include teamwork games, puzzles, and workbased activities.

\section{Discussion}\label{disc}

In this paper, the types of human-AI team currently studied in the literature was explored, to understand what the term means for the field.
To do so, 53 experimental papers exploring human-AI teaming were analysed, revealing strong preferences for teams consisting of 1 human and 1 AI, in the areas of games, object classification, and aviation/military/space.
The studies were categorised using the task and team level characteristic taxonomies provided by \cite{wildman2012task}.
Both humans and AI were most likely to be involved in defined problem-solving tasks, however humans were also more likely to be involved in managing others, whilst AI were more likely to be involved in advising others.
In terms of team level characteristics, five main clusters were observed, representing unique combinations.
These are discussed below.

\subsection{Unique Subtypes of Human-AI Teaming Exist}

The five main clusters found from the analysis are summarised in Table \ref{tab:sum}.
Several insights can be made about the human-AI teaming literature by considering these clusters.
For example, the main areas of interest to the research community can be seen as either falling into Cluster 1 (AI Assistant) or Cluster 2 (Ad hoc Dependency), accounting for 55\% of papers (29/53).
In other words, the majority of studies focus on teaming where an AI assists a human with a task, or a group of at least three team members with different roles working closely together.

\begin{table}[h]
\resizebox{\textwidth}{!}{
\begin{tabular}{@{}llllll@{}}
\toprule
\multicolumn{1}{l}{\textbf{Cluster}} & \textbf{AI Assistant} & \textbf{Ad hoc Dependency} & \textbf{\begin{tabular}[c]{@{}c@{}}Ad hoc Forced\\ Dependency\end{tabular}} & \textbf{\begin{tabular}[c]{@{}c@{}}Paired\\ Equanimity\end{tabular}} & \textbf{\begin{tabular}[c]{@{}c@{}}Group\\ Equanimity\end{tabular}} \\ \midrule
\textbf{Team Makeup} & 2 & 3+ & 2 & 2 & 3+ \\
\textbf{Task Interdependence} & sequential & intensive & reciprocal & reciprocal & intensive \\
\textbf{Role Structure} & functional & functional & functional & divisional & divisional \\
\textbf{Leadership Structure} & designated & distributed & designated & distributed & distributed \\
\textbf{Communication Structure} & chain & star & chain & star & star \\
\textbf{Physical Distribution} & colocated/distributed & colocated/mixed/distributed & colocated & colocated & colocated/mixed \\
\textbf{Team Life Span} & long term/ad hoc & ad hoc/long term & ad hoc/long term & ad hoc & ad hoc/long term \\
\textbf{Most Common Application} & classification & aviation/military/space & games & games & puzzles \\
\textbf{Most Common Task} & object classification & aerialbased & teamwork games & teamwork games & teamwork games \\
\textbf{Number of Studies} & 18 & 11 & 6 & 4 & 3 \\ \bottomrule
\end{tabular}
}
\caption{A summary of the five main clusters of human-AI teaming sub-types.}
\label{tab:sum}
\end{table}

However, we can also see areas that have not received as much attention from the literature.
Considering there were six dimensions of team level characteristics, of which four were used here to cluster studies, the number of potential combinations far surpasses those seen here.
Several papers did not form a cluster as the combination of characteristics was unique.
Paper \cite{islam2025human} for example looked at a team of 1 human with 5 AI drones, where the human provided demonstrations of movement and corrected the AI where appropriate.
Consequently, this represents a team with intensive task interdependence, functional roles, designated leadership, hub-and-wheel communication, colocation, and ad hoc lifespan.
Furthermore, other team types were not seen at all, such as any involving a pooled task interdependence, or an external manager leadership structure.
There is no reason to assume these types of teams cannot exist, or would not be of interest to study for a specific context.
Therefore, the application of the teaming taxonomies to experimental studies can reveal gaps in study, and interesting areas for future research.

\subsection{Implications for Human-AI Teaming Research}

As can be seen, there are strong overlaps in the types of team level characteristics observed between clusters, such as a slight preference for distributed leadership and star communication structures.
However, each cluster is unique in its \textit{combination} of characteristics, creating inherent differences between them.
It is therefore important to consider each one as unique, whilst still falling under the broader umbrella term of `a human-AI team.'

However, doing so presents an issue for the field of human-AI teaming.
Despite their overlaps, each cluster remains unique.
For example, Cluster 1 (AI Assistant) and Cluster 4 (Paired Equanimity) share almost no team level characteristics in common, yet both are `human-AI teams.'
Consequently, any insights gained from studying a human-AI team with the characteristics of Cluster 1 are unlikely to transfer to a study with the team characteristics of Cluster 4.
If, for example, a paper using a Cluster 4 team makes an insight into how to manage leadership roles in a human-AI team (such as \cite{flathmann2024empirically}), it would be difficult to transfer this insight on distributed leadership to Cluster 1, with its designated leadership (such as \cite{mahmood2024designing}).

Currently, authors would be required to read each individual study to assess its similarity to the author's conception of a human-AI team.
There has been a lack of taxonomic language that could aid in this comparison; that is, knowledge that team-level characteristics exist, have been previously defined, and can be used to separate different kinds of teams.
Further, because of this lack of previous language, it may prove challenging for authors to extract this information, depending on the level of information about team level characteristics provided within papers to date.
Alternatively, authors may seek meta reviews which synthesise key insights across papers to create general guidance.
However, the nuances between team types are likely lost in this process, reducing the usefulness and applicability of reviews.
In doing so, authors may find it difficult to understand what guidance applies to their type of human-AI team, and what does not.

Therefore, it is important to not treat all human-AI teaming work as inherently interchangeable, due to differences in team level characteristics.
This is not the same argument as remaining mindful of contexts such as application domain, which has been argued previously (e.g. calls for wider sociotechnical perspectives; \cite{berretta2023defining}).
As seen in the analysis, each cluster contained a range of application domains and task environments, though there were some preferences in certain clusters.
Furthermore, it is important not to consider a team's makeup as the only important distinction between two studies on human-AI teams.
Three of the five main clusters involved 1 human working with 1 AI, however the ways in which they work together vary greatly.
For example, whilst Cluster 3 (Ad hoc Forced Dependency) and Cluster 4 (Paired Equanimity) both involve 1 human and 1 AI, they are not similar in terms of team level characteristics.
As a consequence, the ways in which these humans and AI work together will be markedly different.
Therefore, we argue that care should be taken when synthesising research across studies, so that important nuances are not overlooked.

\subsubsection{Wider Reflections on Applying Teaming Taxonomies to Human-AI Teaming}

The analysis highlights different interpretations of the human-AI teaming definition in experimental work, particularly in terms of interdependence, complementary capabilities, and perceptions of AI as a teammate.
We are not the first to highlight differences in definitions, however others have focused primarily on the \textit{outcomes} of teams.
For example, \cite{lou2025unraveling} considers differences in teaming situational awareness, whilst \cite{mcneese2021my} considers how teaming composition impacts team performance, and \cite{hemmer2025complementarity} considers the impact of complementarity on team performance.
Instead, we focus specifically on the words used when \textit{describing} teams, and what this tells us about how researchers actively interpret the definition.
Each difference in interpretation consequently highlights ambiguities in the definition that require resolution in order to truly synthesise the field, explored below.

Firstly, in human-AI teaming definitions there is an emphasis on interdependence (e.g. \cite{berretta2023defining}).
Similarly, interdependence is also important in team level characteristics taxonomies (e.g. \cite{wildman2012task}), however here the interdependence is further specified into four sub-types (pooled, sequential, reciprocal, and intensive).
Currently, there has not been a discussion in the literature on what interdependence means in practice, and if the type of interdependence is important for human-AI teams.
This highlights a need for a more nuanced conversation around what is meant by interdependence for a given research context.

Secondly, human-AI teaming definitions asks us to consider that humans and AI have ``unique and complementary capabilities'' \cite{berretta2023defining}.
However, as was seen in the results, some team types involve teams where roles are the same between humans and AI (i.e. divisional role structures).
In particular, Clusters 4 and 5 involve divisional roles between humans and AI, seemingly contradicting the requirement that humans and AI have `unique' capabilities.
This implies complementarity is either not an important component of human-AI teaming in all contexts, or certain types of teams studied currently do not align well under the human-AI teaming definition.

The question of whether all experimental studies analysed here align under current conceptions of human-AI teaming is brought into starker contrast for the final requirement that AI is perceived as a teammate.
It is also one of the most important components of the definition, as it separates teaming from other similar terms such as human-AI interaction or human-AI tool use.
By considering the analysis performed here, interpreting this requirement can be challenging.
For example, for Cluster 1, and in some ways Cluster 2, it is difficult to articulate how AI in this team is more than a simple tool.
In Cluster 1, the task level characteristics show that the human remains in control throughout, and is the only one to make a final decision.
Further, the human is technically capable of completing the entire task on their own, with the assumption that AI is there to enhance the performance or efficiency of the task.
Therefore, it is unclear how to describe the AI in these teams as being more than a tool, or conversely, explain what a tool has to be capable of doing before it is no longer considered a tool.
Even considering the team level characteristics cannot aid in distinguishing tools from teammates.
For example, divisional roles may imply a level of similarity between humans and AI, thereby implying a perception of being a teammate.
However, functional roles were very common in experimental studies, and further can indicate that an AI provides a necessary skillset to the team.

Overall, there are issues in applying existing human-AI teaming definitions to practical implementations and concrete examples.
These difficulties stem from two areas.
Firstly, as part of the human-AI teaming definition, we are asked to consider an AI as if it was a human.
This is because the definitions have been built from human-only teaming literature, consequently leaving us to evaluate AI based on human characteristics.
In other words, human-AI teaming, by definition, attempts to ascribe human characteristics to a computer.
This is akin to assigning teaming taxonomies to technologies such as a vehicle, an automatic light, or a calculator.
By taking some generous and interpretive steps, one could consider these technologies to have teaming qualities such as interdependence.
However, is this a \textit{helpful} framing?
Or is the desire to understand human-AI teams within a framework built primarily for human-only teams too restrictive?

Secondly, the wide range of teaming examples observed in this study highlights the need to answer whether the human-AI teaming definition is specific enough for purpose, or in some cases, overused in inappropriate settings.
If it is important that human-AI teams involve AI that are uniquely capable and perceived as teammates, then many of the studies analysed would not fit under this definition.
For example, all of Cluster 1 would be removed for being a simple tool rather than a teammate, and Clusters 4 and 5 would similarly be removed for lack of unique functional roles.
This would remove 25 studies, or 46\% of all studies analysed.
Therefore, the main question for the field becomes: when is a human-AI team a \textit{team}, and what characteristics are truly emblematic of this?

To synthesise and advance the human-AI teaming literature, there is a need to consider more carefully what aspects of teaming should be built into its definition.
In particular, which aspects of teaming are feasible to transfer from human-only to human-AI teams, and what characteristics are unique to human-AI teams, is yet to be determined.
Synthesis is particularly important to achieve as safety-critical applications are increasingly interested in human-AI teaming, particularly due to recent advancements in agentic AI systems and their conversational capabilities.
Clearer taxonomies and definitions are becoming increasingly essential, as they form the basis for three critical processes: (1)~safety analysis and assurance, where defining the scope of the subject of interest is key \cite{habli2025big}; (2)~regulatory frameworks, such as determining how to oversee a collaborative team that goes beyond individually regulated components (e.g. a qualified clinician working with an AI medical device), and; (3)~the allocation of moral responsibility and legal liability \cite{porter2025unravelling}, which mitigates the risk of `responsibility gaps' \cite{matthias2004responsibility} and the `problem of many hands' \cite{royakkers2015moral}. 
However, whilst important to consider, it is unlikely that taxonomies and definitions built solely from human-only teams will be sufficient to capture all nuances of human-AI teams, particularly for safety-critical contexts.
For example, the distinction from tool to teammate, and how to interpret physical distribution, are unlikely resolvable from psychological literature alone.
Therefore, there is a need to develop taxonomies of human-AI teams that build on human-only team literature, but also extends and explains what is unique about having an AI teammate.

\subsection{A Checklist for Reporting Human-AI Teams}

Given the presence of multiple sub-types of human-AI teams present in the literature, it is important that future work more accurately describes what is being studied under the global human-AI teaming definition.
The following checklist is provided as an initial way to help authors consider the specifics of the human-AI teams they are studying.
When designing and reporting human-AI teaming studies, we suggest the following questions be considered, shown in Table \ref{tab:quest}.

\begin{table}[h]
\resizebox{\textwidth}{!}{
\begin{tabular}{@{}ll@{}}
\toprule
\textbf{Question Type} & \textbf{Question} \\ \midrule
General Team Characteristics & How many human and AI team members are in the team? \\
 & What is the intended application domain (e.g. healthcare, aviation, work offices) \\
 & What is the overall team goal? \\
Task Level Characteristics & What roles do human teammates perform? \\
 & What roles do AI teammates perform? \\
 & How do human and AI teammates interact? \\
Team Level Characteristics & What is the task interdependence? (pooled, sequential, reciprocal, intensive) \\
 & What is the role structure? (functional, divisional) \\
 & What is the leadership structure? (external, designated, temporary, distributed) \\
 & What is the communication structure? (hub-and-wheel, chain, star) \\
 & Is physical distribution an important component? \\
 & What is the team life span? (ad hoc, long term) \\ \bottomrule
\end{tabular}
}
\caption{Questions to guide the specification of the type of human-AI team to be studied or reported.}
\label{tab:quest}
\end{table}

In doing so, authors are encouraged to understand their human-AI team more specifically, in turn allowing easier comparison to other work and aiding the identification of overlaps as well as gaps in understanding.
It is not intended to be an exhaustive list, but rather a systematic basis for scoping and framing the conversation around what is unique about human-AI teaming research and its sub-types, and encourage more transparent experimental reporting.
We also invite authors to consider what other questions are important to consider when describing a human-AI team, to build upon this initial step.

\subsection{Limitations \& Future Work}

There are a few limitations to the work undertaken in this study.
Firstly, only papers that explicitly included the term ``human-AI team*" in the title or keywords were included in the sample.
It is possible papers were missed that did not include this specific phrase (such as human agent teaming, or human autonomy teaming), or did not do so in the title or keywords.
However, a large number of papers were still collected, and can still provide a scoping overview of the current experimental literature.

Secondly, an established taxonomy was used to conduct the content analysis, to reduce the risk of bias in interpreting teaming characteristics.
However, the taxonomy required slight interpretations to apply to human-AI teams, which could have introduced errors in coding.
To enhance transparency, we have included the dataset of all papers analysed, as well as their assigned categories.
We encourage readers to review the data and consider how the taxonomic labels apply to the papers in the set, as well as other literature in the field.

In terms of future work, the current study highlighted the existence of multiple sub-types of human-AI teams, all studied under the same term.
Some combinations of team characteristics are more commonly studied than others, indicating potential areas of future research in these understudied teams.
For example, future work could look into characteristics such as pooled task interdependence or external manager leadership, and how these influence human-AI teaming.

However, it is also important to note the sub-types are unique and not interchangeable.
Consequently, there is a need for more clarity in how human-AI teams are described in future research.
We encourage readers to reflect on what types of human-AI teams they are interested in, and what characteristics of the team are important for fellow readers to know in order to understand the team.
To this end, to aid in effective research communication, we also encourage readers to use the proposed checklist as a baseline for how to report a human-AI team in experimental work.

\section{Conclusion}\label{sec13}

To further advance the field of human-AI teaming, there is a need for more precise language to describe the types of teams that are studied.
We propose an initial checklist on key features to report and consider when designing a team, and encourage authors to reflect more broadly on what types of teaming they are interested in.
In doing so, we hope there will be a better synthesis of human-AI teaming literature, as well as new avenues in exploring currently understudied combinations of team level characteristics.

\section{Acknowledgements}

This work was supported by the Centre for Assuring Autonomy, a partnership between Lloyd’s Register Foundation and the University of York.

\bibliography{bib}

@book{royakkers2015moral,
  title={Moral responsibility and the problem of many hands},
  author={Royakkers, Lamb{\`e}r and Sjoerd D.. Zwart},
  year={2015},
  publisher={Routledge}
}

@article{matthias2004responsibility,
  title={The responsibility gap: Ascribing responsibility for the actions of learning automata},
  author={Matthias, Andreas},
  journal={Ethics and information technology},
  volume={6},
  number={3},
  pages={175--183},
  year={2004},
  publisher={Springer}
}

@article{porter2025unravelling,
  title={Unravelling responsibility for AI},
  author={Porter, Zoe and Ryan, Philippa and Morgan, Phillip and Al-Qaddoumi, Joanna and Twomey, Bernard and Noordhof, Paul and McDermid, John and Habli, Ibrahim},
  journal={Journal of Responsible Technology},
  pages={100124},
  year={2025},
  publisher={Elsevier}
}

@article{habli2025big,
  title={The big argument for ai safety cases},
  author={Habli, Ibrahim and Hawkins, Richard and Paterson, Colin and Ryan, Philippa and Jia, Yan and Sujan, Mark and McDermid, John},
  journal={arXiv preprint arXiv:2503.11705},
  year={2025}
}

@article{wildman2012task,
  title={Task types and team-level attributes: Synthesis of team classification literature},
  author={Wildman, Jessica L and Thayer, Amanda L and Rosen, Michael A and Salas, Eduardo and Mathieu, John E and Rayne, Sara R},
  journal={Human resource development review},
  volume={11},
  number={1},
  pages={97--129},
  year={2012},
  publisher={Sage Publications Sage CA: Los Angeles, CA}
}

@article{munn2018systematic,
  title={Systematic review or scoping review? Guidance for authors when choosing between a systematic or scoping review approach},
  author={Munn, Zachary and Peters, Micah DJ and Stern, Cindy and Tufanaru, Catalin and McArthur, Alexa and Aromataris, Edoardo},
  journal={BMC medical research methodology},
  volume={18},
  pages={1--7},
  year={2018},
  publisher={Springer}
}

@article{gao2023agent,
  title={Agent teaming situation awareness (ATSA): A situation awareness framework for human-AI teaming},
  author={Gao, Qi and Xu, Wei and Shen, Mowei and Gao, Zaifeng},
  journal={arXiv preprint arXiv:2308.16785},
  year={2023}
}

@article{mcgrath2025collaborative,
  title={Collaborative human-AI trust (CHAI-T): A process framework for active management of trust in human-AI collaboration},
  author={McGrath, Melanie J and Duenser, Andreas and Lacey, Justine and Paris, Cecile},
  journal={Computers in Human Behavior: Artificial Humans},
  pages={100200},
  year={2025},
  publisher={Elsevier}
}

@article{arksey2005scoping,
  title={Scoping studies: towards a methodological framework},
  author={Arksey, Hilary and O'malley, Lisa},
  journal={International journal of social research methodology},
  volume={8},
  number={1},
  pages={19--32},
  year={2005},
  publisher={Taylor \& Francis}
}

@book{krippendorff2018content,
  title={Content analysis: An introduction to its methodology},
  author={Krippendorff, Klaus},
  year={2018},
  publisher={Sage publications}
}

@article{tricco2018prisma,
  title={PRISMA extension for scoping reviews (PRISMA-ScR): checklist and explanation},
  author={Tricco, Andrea C and Lillie, Erin and Zarin, Wasifa and O'Brien, Kelly K and Colquhoun, Heather and Levac, Danielle and Moher, David and Peters, Micah DJ and Horsley, Tanya and Weeks, Laura and others},
  journal={Annals of internal medicine},
  volume={169},
  number={7},
  pages={467--473},
  year={2018},
  publisher={American College of Physicians}
}

@article{berretta2023defining,
  title={Defining human-AI teaming the human-centered way: a scoping review and network analysis},
  author={Berretta, Sophie and Tausch, Alina and Ontrup, Greta and Gilles, Bj{\"o}rn and Peifer, Corinna and Kluge, Annette},
  journal={Frontiers in Artificial Intelligence},
  volume={6},
  pages={1250725},
  year={2023},
  publisher={Frontiers Media SA}
}

@article{lou2025unraveling,
  title={Unraveling human-AI teaming: a review and outlook},
  author={Lou, Bowen and Lu, Tian and Raghu, TS and Zhang, Yingjie},
  journal={arXiv preprint arXiv:2504.05755},
  year={2025}
}

@article{mcneese2021my,
  title={Who/what is my teammate? Team composition considerations in human--AI teaming},
  author={McNeese, Nathan J and Schelble, Beau G and Canonico, Lorenzo Barberis and Demir, Mustafa},
  journal={IEEE Transactions on Human-Machine Systems},
  volume={51},
  number={4},
  pages={288--299},
  year={2021},
  publisher={IEEE}
}

@article{hemmer2025complementarity,
  title={Complementarity in human-AI collaboration: Concept, sources, and evidence},
  author={Hemmer, Patrick and Schemmer, Max and K{\"u}hl, Niklas and V{\"o}ssing, Michael and Satzger, Gerhard},
  journal={European Journal of Information Systems},
  volume={34},
  number={6},
  pages={979--1002},
  year={2025},
  publisher={Taylor \& Francis}
}

@article{bansal2019case,
  title={A case for backward compatibility for human-ai teams},
  author={Bansal, Gagan and Nushi, Besmira and Kamar, Ece and Weld, Dan and Lasecki, Walter and Horvitz, Eric},
  journal={arXiv preprint arXiv:1906.01148},
  year={2019}
}

@article{olla2024cybersecurity,
  title={A cybersecurity game to probe human-AI teaming},
  author={Olla, Rita and Hand, Emily and Louis, Sushil J and Houmanfar, Ramona and Sengupta, Shamik},
  journal={2024 IEEE Conference on Games (CoG)},
  pages={p. 1--5},
  year={2024},
}

@article{amresh2023minecraft,
  title={A Minecraft based simulated task environment for human AI teaming},
  author={Amresh, Ashish and Cooke, Nancy and Fouse, Adam},
  journal={Proceedings of the 23rd ACM international conference on intelligent virtual agents},
  pages={p. 1--3},
  year={2023}
}

@article{schwalb2022study,
  title={A study of drone-based AI for enhanced human-AI trust and informed decision making in human-AI interactive virtual environments},
  author={Schwalb, Joseph and Menon, Vineetha and Tenhundfeld, Nathan and Weger, Kristin and Mesmer, Bryan and Gholston, Sampson},
  journal={2022 IEEE 3rd International Conference on Human-Machine Systems (ICHMS)},
  pages={p. 1--6},
  year={2022},
}

@article{tariq2025a2c,
  title={A2C: A modular multi-stage collaborative decision framework for human--AI teams},
  author={Tariq, Shahroz and Chhetri, Mohan Baruwal and Nepal, Surya and Paris, Cecile},
  journal={Expert systems with applications},
  volume={282},
  pages={127318},
  year={2025},
  publisher={Elsevier}
}

@article{josephs2023artifact,
  title={Artifact magnification on deepfake videos increases human detection and subjective confidence},
  author={Josephs, Emilie and Fosco, Camilo and Oliva, Aude},
  journal={arXiv preprint arXiv:2304.04733},
  year={2023}
}

@article{ong2012closing,
  title={Closing the human-AI team-mate gap: how changes to displayed information impact player behavior towards computer teammates},
  author={Ong, Christopher and McGee, Kevin and Chuah, Teong Leong},
  journal={Proceedings of the 24th Australian Computer-Human Interaction Conference},
  pages={p. 433--439},
  year={2012}
}

@article{zvelebilova2024collective,
  title={Collective attention in human-AI teams},
  author={Zvelebilova, Josie and Savage, Saiph and Riedl, Christoph},
  journal={arXiv preprint arXiv:2407.17489},
  year={2024}
}

@article{xu2023comparing,
  title={Comparing zealous and restrained ai recommendations in a real-world human-ai collaboration task},
  author={Xu, Chengyuan and Lien, Kuo-Chin and H{\"o}llerer, Tobias},
  journal={Proceedings of the 2023 CHI Conference on Human Factors in Computing Systems},
  pages={p. 1--15},
  year={2023}
}

@article{zhang2024crew,
  title={Crew: Facilitating human-ai teaming research},
  author={Zhang, Lingyu and Ji, Zhengran and Chen, Boyuan},
  journal={arXiv preprint arXiv:2408.00170},
  year={2024}
}

@article{munyaka2023decision,
  title={Decision making strategies and team efficacy in human-AI teams},
  author={Munyaka, Imani and Ashktorab, Zahra and Dugan, Casey and Johnson, James and Pan, Qian},
  journal={Proceedings of the ACM on Human-Computer Interaction},
  volume={7},
  number={CSCW1},
  pages={1--24},
  year={2023},
  publisher={ACM New York, NY, USA}
}

@article{mahmood2024designing,
  title={Designing behavior-aware AI to improve the human-AI team performance in AI-assisted decision making},
  author={Mahmood, Syed\_Hasan Amin and Lu, Zhuoran and Yin, Ming},
  year={2024},
  journal={International Joint Conferences on Artificial Intelligence Organization}
}

@article{newn2019designing,
  title={Designing interactions with intention-aware gaze-enabled artificial agents},
  author={Newn, Joshua and Singh, Ronal and Allison, Fraser and Madumal, Prashan and Velloso, Eduardo and Vetere, Frank},
  journal={IFIP Conference on Human-Computer Interaction},
  pages={p. 255--281},
  year={2019}
}

@article{bansal2021does,
  title={Does the whole exceed its parts? the effect of ai explanations on complementary team performance},
  author={Bansal, Gagan and Wu, Tongshuang and Zhou, Joyce and Fok, Raymond and Nushi, Besmira and Kamar, Ece and Ribeiro, Marco Tulio and Weld, Daniel},
  journal={Proceedings of the 2021 CHI conference on human factors in computing systems},
  pages={p. 1--16},
  year={2021}
}

@article{flathmann2024empirically,
  title={Empirically understanding the potential impacts and process of social influence in human-AI teams},
  author={Flathmann, Christopher and Duan, Wen and Mcneese, Nathan J and Hauptman, Allyson and Zhang, Rui},
  journal={Proceedings of the ACM on Human-Computer Interaction},
  volume={8},
  number={CSCW1},
  pages={1--32},
  year={2024},
  publisher={ACM New York, NY, USA}
}

@article{flathmann2023examining,
  title={Examining the impact of varying levels of AI teammate influence on human-AI teams},
  author={Flathmann, Christopher and Schelble, Beau G and Rosopa, Patrick J and McNeese, Nathan J and Mallick, Rohit and Madathil, Kapil Chalil},
  journal={International Journal of Human-Computer Studies},
  volume={177},
  pages={103061},
  year={2023},
  publisher={Elsevier}
}

@article{jorge2024should,
  title={How should an AI trust its human teammates? Exploring possible cues of artificial trust},
  author={Jorge, Carolina Centeio and Jonker, Catholijn M and Tielman, Myrthe L},
  journal={ACM Transactions on Interactive Intelligent Systems},
  volume={14},
  number={1},
  pages={1--26},
  year={2024},
  publisher={ACM New York, NY}
}

@article{sidji2024human,
  title={Human-AI collaboration in cooperative games: A study of playing codenames with an LLM assistant},
  author={Sidji, Matthew and Smith, Wally and Rogerson, Melissa J},
  journal={Proceedings of the ACM on Human-Computer Interaction},
  volume={8},
  number={CHI PLAY},
  pages={1--25},
  year={2024},
  publisher={ACM New York, NY, USA}
}

@article{islam2025human,
  title={Human-AI collaboration in real-world complex environment with reinforcement learning},
  author={Islam, Md Saiful and Das, Srijita and Gottipati, Sai Krishna and Duguay, William and Mars, Clodric and Arabneydi, Jalal and Fagette, Antoine and Guzdial, Matthew and Taylor, Matthew E},
  journal={Neural Computing and Applications},
  volume={37},
  number={23},
  pages={18957--18987},
  year={2025},
  publisher={Springer}
}

@article{bienefeld2023human,
  title={Human-AI teaming: leveraging transactive memory and speaking up for enhanced team effectiveness},
  author={Bienefeld, Nadine and Kolbe, Michaela and Camen, Giovanni and Huser, Dominic and Buehler, Philipp Karl},
  journal={Frontiers in Psychology},
  volume={14},
  pages={1208019},
  year={2023},
  publisher={Frontiers Media SA}
}

@article{momose2025human,
  title={Human-AI teamwork interface design using patterns of interactions},
  author={Momose, Kazuhiko and Mehta, Rahul and Moukpe, Josias and Weekes, Troy R and Eskridge, Thomas C},
  journal={International Journal of Human--Computer Interaction},
  volume={41},
  number={11},
  pages={7112--7134},
  year={2025},
  publisher={Taylor \& Francis}
}

@article{schadd2022m,
  title={“i’m afraid i can’t do that, dave”; getting to know your buddies in a human--agent team},
  author={Schadd, Maarten PD and Schoonderwoerd, Tjeerd AJ and van den Bosch, Karel and Visker, Olaf H and Haije, Tjalling},
  journal={Systems},
  volume={10},
  number={1},
  pages={15},
  year={2022},
  publisher={MDPI}
}

@article{bhambri2023incorporating,
  title={Incorporating human flexibility through reward preferences in human-AI teaming},
  author={Bhambri, Siddhant and Verma, Mudit and Biswas, Upasana and Murthy, Anil and Kambhampati, Subbarao},
  journal={arXiv preprint arXiv:2312.14292},
  year={2023}
}

@article{zhang2023investigating,
  title={Investigating AI teammate communication strategies and their impact in human-AI teams for effective teamwork},
  author={Zhang, Rui and Duan, Wen and Flathmann, Christopher and McNeese, Nathan and Freeman, Guo and Williams, Alyssa},
  journal={Proceedings of the ACM on Human-Computer Interaction},
  volume={7},
  number={CSCW2},
  pages={1--31},
  year={2023},
  publisher={ACM New York, NY, USA}
}

@article{ye2022modeling,
  title={Modeling human-ai team decision making},
  author={Ye, Wei and Bullo, Francesco and Friedkin, Noah and Singh, Ambuj K},
  journal={arXiv preprint arXiv:2201.02759},
  year={2022}
}

@article{li2024modeling,
  title={Modeling trust dimensions and dynamics in human-agent conversation: A trajectory epistemic network analysis approach},
  author={Li, Mengyao and Kamaraj, Amudha V and Lee, John D},
  journal={International Journal of Human--Computer Interaction},
  volume={40},
  number={14},
  pages={3571--3582},
  year={2024},
  publisher={Taylor \& Francis}
}

@article{attig2024more,
  title={More than task performance: Developing new criteria for successful human-AI teaming using the cooperative card game Hanabi},
  author={Attig, Christiane and Wollstadt, Patricia and Schrills, Tim and Franke, Thomas and Wiebel-Herboth, Christiane B},
  journal={Extended abstracts of the chi conference on human factors in computing systems},
  pages={p. 1--11},
  year={2024}
}

@article{zhang2024mutual,
  title={Mutual theory of mind in human-ai collaboration: An empirical study with llm-driven ai agents in a real-time shared workspace task},
  author={Zhang, Shao and Wang, Xihuai and Zhang, Wenhao and Chen, Yongshan and Gao, Landi and Wang, Dakuo and Zhang, Weinan and Wang, Xinbing and Wen, Ying},
  journal={arXiv preprint arXiv:2409.08811},
  year={2024}
}

@article{smith2025navigating,
  title={Navigating AI convergence in human--artificial intelligence teams: A signaling theory approach},
  author={Smith, Andria and van Wagoner, Hunter Phoenix and Keplinger, Ksenia and Celebi, Can},
  journal={Journal of Organizational Behavior},
  year={2025},
  publisher={Wiley Online Library}
}

@article{darban2024navigating,
  title={Navigating virtual teams in generative AI-led learning: The moderation of team perceived virtuality},
  author={Darban, Mehdi},
  journal={Education and Information Technologies},
  volume={29},
  number={17},
  pages={23225--23248},
  year={2024},
  publisher={Springer}
}

@article{babbar2022utility,
  title={On the utility of prediction sets in human-ai teams},
  author={Babbar, Varun and Bhatt, Umang and Weller, Adrian},
  journal={arXiv preprint arXiv:2205.01411},
  year={2022}
}

@article{wurfel2024operationalizing,
  title={Operationalizing ai explainability using interpretability cues in the cockpit: Insights from user-centered development of the intelligent pilot advisory system (IPAS)},
  author={W{\"u}rfel, Jakob and Papenfu{\ss}, Anne and Wies, Matthias},
  journal={International Conference on Human-Computer Interaction},
  pages={p. 297--315},
  year={2024}
}

@article{lu2023readingquizmaker,
  author={Lu, Xinyi and Fan, Simin and Houghton, Jessica and Wang, Lu and Wang, Xu},
  title={ReadingQuizMaker: a human-NLP collaborative system that supports instructors to design high-quality reading quiz questions},
  journal={Proceedings of the 2023 CHI Conference on Human Factors in Computing Systems},
  pages={{p. 1--18}},
  year={2023}
}

@article{prieto2023single,
  title={Single-case learning analytics: Feasibility of a human-centered analytics approach to support doctoral education},
  author={Prieto Santos, Luis Pablo and Pishtari, Gerti and Dimitriadis, Yannis and Rodr{\'\i}guez Triana, Mar{\'\i}a Jes{\'u}s and Ley, Tobias and Odriozola Gonz{\'a}lez, Paula and others},
  journal={JUCS-Journal of Universal Computer Science},
  volume={29},
  number={9},
  pages={1033--1068},
  year={2023},
  publisher={Graz University of Technology}
}

@article{harris2023social,
  title={Social perception in Human-AI teams: Warmth and competence predict receptivity to AI teammates},
  author={Harris-Watson, Alexandra M and Larson, Lindsay E and Lauharatanahirun, Nina and DeChurch, Leslie A and Contractor, Noshir S},
  journal={Computers in Human Behavior},
  volume={145},
  pages={107765},
  year={2023},
  publisher={Elsevier}
}

@article{milella2023impact,
  title={The impact of gender and personality in human-AI teaming: The case of collaborative question answering},
  author={Milella, Frida and Natali, Chiara and Scantamburlo, Teresa and Campagner, Andrea and Cabitza, Federico},
  journal={IFIP Conference on Human-Computer Interaction},
  pages={p. 329--349},
  year={2023}
}

@article{mallick2024pursuit,
  title={The pursuit of happiness: the power and influence of AI teammate emotion in human-AI teamwork},
  author={Mallick, Rohit and Flathmann, Christopher and Lancaster, Caitlin and Hauptman, Allyson and McNeese, Nathan and Freeman, Guo},
  journal={Behaviour \& Information Technology},
  volume={43},
  number={14},
  pages={3436--3460},
  year={2024},
  publisher={Taylor \& Francis}
}

@article{zhao2025role,
  title={The role of adaptation in collective human--AI teaming},
  author={Zhao, Michelle and Simmons, Reid and Admoni, Henny},
  journal={Topics in cognitive science},
  volume={17},
  number={2},
  pages={291--323},
  year={2025},
  publisher={Wiley Online Library}
}

@article{schelble2024towards,
  title={Towards ethical AI: Empirically investigating dimensions of AI ethics, trust repair, and performance in human-AI teaming},
  author={Schelble, Beau G and Lopez, Jeremy and Textor, Claire and Zhang, Rui and McNeese, Nathan J and Pak, Richard and Freeman, Guo},
  journal={Human Factors},
  volume={66},
  number={4},
  pages={1037--1055},
  year={2024},
  publisher={Sage Publications Sage CA: Los Angeles, CA}
}

@article{jia2021towards,
  title={Towards visual explainable active learning for zero-shot classification},
  author={Jia, Shichao and Li, Zeyu and Chen, Nuo and Zhang, Jiawan},
  journal={IEEE Transactions on Visualization and Computer Graphics},
  volume={28},
  number={1},
  pages={791--801},
  year={2021},
  publisher={IEEE}
}

@article{marrone2025understanding,
  title={Understanding student perceptions of artificial intelligence as a teammate},
  author={Marrone, Rebecca and Zamecnik, Andrew and Joksimovic, Srecko and Johnson, Jarrod and De Laat, Maarten},
  journal={Technology, Knowledge and Learning},
  volume={30},
  number={3},
  pages={1847--1869},
  year={2025},
  publisher={Springer}
}

@article{duan2024understanding,
  title={Understanding the evolvement of trust over time within Human-AI teams},
  author={Duan, Wen and Zhou, Shiwen and Scalia, Matthew J and Yin, Xiaoyun and Weng, Nan and Zhang, Ruihao and Freeman, Guo and McNeese, Nathan and Gorman, Jamie and Tolston, Michael},
  journal={Proceedings of the ACM on Human-Computer Interaction},
  volume={8},
  number={CSCW2},
  pages={1--31},
  year={2024},
  publisher={ACM New York, NY, USA}
}

@article{hauptman2024understanding,
  title={Understanding the influence of AI autonomy on AI explainability levels in human-AI teams using a mixed methods approach},
  author={Hauptman, Allyson I and Schelble, Beau G and Duan, Wen and Flathmann, Christopher and McNeese, Nathan J},
  journal={Cognition, Technology \& Work},
  volume={26},
  number={3},
  pages={435--455},
  year={2024},
  publisher={Springer}
}

@article{duan2025understanding,
  title={Understanding the processes of trust and distrust contagion in human--AI teams: A qualitative approach},
  author={Duan, Wen and Zhou, Shiwen and Scalia, Matthew J and Freeman, Guo and Gorman, Jamie and Tolston, Michael and McNeese, Nathan J and Funke, Gregory},
  journal={Computers in Human Behavior},
  volume={165},
  pages={108560},
  year={2025},
  publisher={Elsevier}
}

@article{bansal2019updates,
  title={Updates in human-ai teams: Understanding and addressing the performance/compatibility tradeoff},
  author={Bansal, Gagan and Nushi, Besmira and Kamar, Ece and Weld, Daniel S and Lasecki, Walter S and Horvitz, Eric},
  journal={Proceedings of the AAAI conference on artificial intelligence},
  volume={33},
  number={01},
  pages={p. 2429--2437},
  year={2019}
}

@article{zhang2024verbal,
  title={Verbal vs. visual: How humans perceive and collaborate with AI teammates using different communication modalities in various human-AI team compositions},
  author={Zhang, Rui and Duan, Wen and Flathmann, Christopher and McNeese, Nathan and Knijnenburg, Bart and Freeman, Guo},
  journal={Proceedings of the ACM on human-computer Interaction},
  volume={8},
  number={CSCW2},
  pages={1--34},
  year={2024},
  publisher={ACM New York, NY, USA}
}

@article{hong2023visualizing,
  title={Visualizing and comparing machine learning predictions to improve Human-AI teaming on the example of cell lineage},
  author={Hong, Jiayi and Maciejewski, Ross and Trubuil, Alain and Isenberg, Tobias},
  journal={IEEE Transactions on Visualization and Computer Graphics},
  volume={30},
  number={4},
  pages={1956--1969},
  year={2023},
  publisher={IEEE}
}

@article{mallick2024you,
  title={What you say vs what you do: Utilizing positive emotional expressions to relay AI teammate intent within human--AI teams},
  author={Mallick, Rohit and Flathmann, Christopher and Duan, Wen and Schelble, Beau G and McNeese, Nathan J},
  journal={International Journal of Human-Computer Studies},
  volume={192},
  pages={103355},
  year={2024},
  publisher={Elsevier}
}

@article{georganta2024would,
  title={Would you trust an AI team member? Team trust in human--AI teams},
  author={Georganta, Eleni and Ulfert, Anna-Sophie},
  journal={Journal of occupational and organizational psychology},
  volume={97},
  number={3},
  pages={1212--1241},
  year={2024},
  publisher={Wiley Online Library}
}

@article{zhang2022you,
  title={You complete me: Human-ai teams and complementary expertise},
  author={Zhang, Qiaoning and Lee, Matthew L and Carter, Scott},
  journal={Proceedings of the 2022 CHI conference on human factors in computing systems},
  pages={p. 1--28},
  year={2022}
}

@article{erengin2024you,
  title={You, Me, and the AI: The Role of Third-Party Human Teammates for Trust Formation Toward AI Teammates},
  author={Erengin, T{\"u}rk{\"u} and Briker, Roman and de Jong, Simon B},
  journal={Journal of Organizational Behavior},
  year={2024},
  publisher={Wiley Online Library}
}

@article{wang2020human,
  title={From human-human collaboration to Human-AI collaboration: Designing AI systems that can work together with people},
  author={Wang, Dakuo and Churchill, Elizabeth and Maes, Pattie and Fan, Xiangmin and Shneiderman, Ben and Shi, Yuanchun and Wang, Qianying},
  journal={Extended abstracts of the 2020 CHI conference on human factors in computing systems},
  pages={p. 1--6},
  year={2020}
}

@article{vcartolovni2022ethical,
  title={Ethical, legal, and social considerations of AI-based medical decision-support tools: a scoping review},
  author={{\v{C}}artolovni, Anto and Tomi{\v{c}}i{\'c}, Ana and Mosler, Elvira Lazi{\'c}},
  journal={International Journal of Medical Informatics},
  volume={161},
  pages={104738},
  year={2022},
  publisher={Elsevier}
}

@article{o2022human,
  title={Human--autonomy teaming: A review and analysis of the empirical literature},
  author={O’neill, Thomas and McNeese, Nathan and Barron, Amy and Schelble, Beau},
  journal={Human factors},
  volume={64},
  number={5},
  pages={904--938},
  year={2022},
  publisher={Sage Publications Sage CA: Los Angeles, CA}
}

@article{mcneese2018teaming,
  title={Teaming with a synthetic teammate: Insights into human-autonomy teaming},
  author={McNeese, Nathan J and Demir, Mustafa and Cooke, Nancy J and Myers, Christopher},
  journal={Human factors},
  volume={60},
  number={2},
  pages={262--273},
  year={2018},
  publisher={Sage Publications Sage CA: Los Angeles, CA}
}

@article{salas2008teams,
  title={On teams, teamwork, and team performance: Discoveries and developments},
  author={Salas, Eduardo and Cooke, Nancy J and Rosen, Michael A},
  journal={Human factors},
  volume={50},
  number={3},
  pages={540--547},
  year={2008},
  publisher={SAGE Publications Sage CA: Los Angeles, CA}
}

@article{zhang2021ideal,
  title={" An ideal human" expectations of AI teammates in human-AI teaming},
  author={Zhang, Rui and McNeese, Nathan J and Freeman, Guo and Musick, Geoff},
  journal={Proceedings of the ACM on Human-Computer Interaction},
  volume={4},
  number={CSCW3},
  pages={1--25},
  year={2021},
  publisher={ACM New York, NY, USA}
}

@article{loper2023evolving,
  title={Evolving Lvc To Include Evaluation Of Human-Ai Teaming Dynamics},
  author={Loper, Margaret and Sitterle, Valerie},
  journal={2023 Winter Simulation Conference (WSC)},
  pages={p. 2506--2517},
  year={2023}
}

@article{cabour2023explanation,
  title={An explanation space to align user studies with the technical development of Explainable AI},
  author={Cabour, Garrick and Morales-Forero, Andr{\'e}s and Ledoux, {\'E}lise and Bassetto, Samuel},
  journal={AI \& SOCIETY},
  volume={38},
  number={2},
  pages={869--887},
  year={2023},
  publisher={Springer}
}

@article{flathmann2021modeling,
  title={Modeling and guiding the creation of ethical human-AI teams},
  author={Flathmann, Christopher and Schelble, Beau G and Zhang, Rui and McNeese, Nathan J},
  journal={Proceedings of the 2021 AAAI/ACM Conference on AI, Ethics, and Society},
  pages={p. 469--479},
  year={2021}
}

\clearpage

\section{Appendix}
\begin{table}[h]
\resizebox{\textwidth}{!}{
\begin{tabular}{@{}lllllllllll@{}}
\toprule
\textbf{Cluster} & \textbf{Paper} & \textbf{Application Type} & \textbf{Task Type} & \textbf{Team Makeup} & \textbf{Task interdependence} & \textbf{Role structure} & \textbf{Leadership structure} & \textbf{Communication structure} & \textbf{Physical distribution} & \textbf{Team life span} \\ \midrule
1 & \cite{zhang2024verbal} & Generic/Theoretical & Teamwork Games & 1 human 2 AI; 2 human 1 AI & Intensive & Divisional & Distributed & Star & Colocated & Ad hoc \\
1 & \cite{milella2023impact} & Generic/Theoretical & Quiz & 3-4 humans 1 AI & Intensive & Divisional & Distributed & Star & Mixed & Ad hoc \\
1 & \cite{darban2024navigating} & Education & Workbased Activity & 6-8 humans 1 AI & Intensive & Divisional & Distributed & Star & Mixed & Long Term \\ \midrule
2 & \cite{harris2023social} & Misc. Teamwork & Workbased Activity & 2-3 humans 1 AI & Intensive & Divisional & Temporary & Star & Distributed & Ad hoc \\ \midrule
3 & \cite{islam2025human} & Aviation/Military/Space & Aerialbased & 1 human 5 AI & Intensive & Functional & Designated & Hub-and-Wheel & Colocated & Ad hoc \\ \midrule
4 & \cite{munyaka2023decision} & Game & Teamwork Games & 1 human 2 AI & Intensive & Functional & Distributed & Hub-and-Wheel & Colocated & Ad hoc \\ \midrule
5 & \cite{mallick2024pursuit} & Game & Teamwork Games & 1 human 5-7 AI & Intensive & Functional & Distributed & Star & Colocated & Ad hoc \\
5 & \cite{mallick2024you} & Generic/Theoretical & Teamwork Games & 1 human 5-7 AI & Intensive & Functional & Distributed & Star & Colocated & Ad hoc \\
5 & \cite{duan2024understanding} & Aviation/Military/Space & Aerialbased & 1 human 2 AI; 2 human 1 AI & Intensive & Functional & Distributed & Star & Distributed & Ad hoc \\
5 & \cite{duan2025understanding} & Aviation/Military/Space & Aerialbased & 1 human 2 AI; 2 human 1 AI & Intensive & Functional & Distributed & Star & Distributed & Ad hoc \\
5 & \cite{bienefeld2023human} & Healthcare & Workbased Activity & 4 humans 1 AI & Intensive & Functional & Distributed & Star & Colocated & Ad hoc \\
5 & \cite{marrone2025understanding} & Education & Space & 2-5 humans 1 AI & Intensive & Functional & Distributed & Star & Colocated & Ad hoc \\
5 & \cite{erengin2024you} & Cybersecurity/Security & Security & 2 humans 1 AI & Intensive & Functional & Distributed & Star & Colocated & Long Term \\
5 & \cite{schelble2024towards} & Aviation/Military/Space & Aerialbased & 2 humans 1 AI & Intensive & Functional & Distributed & Star & Distributed & Ad hoc \\
5 & \cite{mcneese2021my} & Disaster Response & Search and Rescue & 1 human 2 AI; 2 human 1 AI & Intensive & Functional & Distributed & Star & Distributed & Ad hoc \\
5 & \cite{amresh2023minecraft} & Disaster Response & Search and Rescue & 3 humans 1 AI & Intensive & Functional & Distributed & Star & Mixed & Ad hoc \\
5 & \cite{zvelebilova2024collective} & Generic/Theoretical & Teamwork Games & 3-4 humans 1 AI & Intensive & Functional & Distributed & Star & Mixed & Ad hoc \\ \midrule
6 & \cite{wurfel2024operationalizing} & Aviation/Military/Space & Aerialbased & 2 humans 1 AI & Intensive & Functional & Temporary & Star & Colocated & Ad hoc \\ \midrule
7 & \cite{li2024modeling} & Aviation/Military/Space & Space & 1 human 1 AI & Reciprocal & Divisional & Designated & Chain & Colocated & Ad hoc \\
7 & \cite{lu2023readingquizmaker} & Education & Workbased Activity & 1 human 1 AI & Reciprocal & Divisional & Designated & Chain & Colocated & Long Term \\ \midrule
8 & \cite{momose2025human} & Aviation/Military/Space & Space & 1 human 1 AI & Reciprocal & Divisional & Distributed & Star & Colocated & Ad hoc \\
8 & \cite{flathmann2024empirically} & Game & Teamwork Games & 1 human 1 AI & Reciprocal & Divisional & Distributed & Star & Colocated & Ad hoc \\
8 & \cite{flathmann2023examining} & Game & Teamwork Games & 1 human 1 AI & Reciprocal & Divisional & Distributed & Star & Colocated & Ad hoc \\
8 & \cite{zhang2023investigating} & Game & Teamwork Games & 1 human 1 AI & Reciprocal & Divisional & Distributed & Star & Colocated & Ad hoc \\ \midrule
9 & \cite{schwalb2022study} & Disaster Response & Search and Rescue & 1 human 1 AI & Reciprocal & Functional & Designated & Chain & Colocated & Ad hoc \\
9 & \cite{newn2019designing} & Game & Teamwork Games & 1 human 1 AI & Reciprocal & Functional & Designated & Chain & Colocated & Ad hoc \\
9 & \cite{bhambri2023incorporating} & Game & Teamwork Games & 1 human 1 AI & Reciprocal & Functional & Designated & Chain & Colocated & Ad hoc \\
9 & \cite{zhang2024mutual} & Work & Teamwork Games & 1 human 1 AI & Reciprocal & Functional & Designated & Chain & Colocated & Ad hoc \\
9 & \cite{hauptman2024understanding} & Education & Workbased Activity & 1 human 1 AI & Reciprocal & Functional & Designated & Chain & Colocated & Ad hoc \\
9 & \cite{tariq2025a2c} & Cybersecurity/Security & Security & 1 human 1 AI & Reciprocal & Functional & Designated & Chain & Colocated & Long Term \\ \midrule
10 & \cite{schadd2022m} & Disaster Response & Search and Rescue & 1 human 1 AI & Reciprocal & Functional & Distributed & Star & Colocated & Ad hoc \\
10 & \cite{georganta2024would} & Work & Workbased Activity & 1 human 1 AI & Reciprocal & Functional & Distributed & Star & Distributed & Long Term \\ \midrule
11 & \cite{sidji2024human} & Game & Teamwork Games & 1 human 1 AI & Sequential & Divisional & Designated & Chain & Colocated & Ad hoc \\ \midrule
12 & \cite{attig2024more} & Game & Teamwork Games & 1 human 1 AI & Sequential & Divisional & Temporary & Star & Colocated & Ad hoc \\ \midrule
13 & \cite{ong2012closing} & Game & Teamwork Games & 1 human 1 AI & Sequential & Functional & Designated & Chain & Colocated & Ad hoc \\
13 & \cite{prieto2023single} & Education & Workbased Activity & 1 human 1 AI & Sequential & Functional & Designated & Chain & Colocated & Long Term \\
13 & \cite{zhang2024crew} & Experimental Game & Teamwork Games & 1 human 1 AI & Sequential & Functional & Designated & Chain & Colocated & Ad hoc \\
13 & \cite{xu2023comparing} & Classification & Object Classification & 1 human 1 AI & Sequential & Functional & Designated & Chain & Colocated & Long Term \\
13 & \cite{hemmer2025complementarity} & Classification & Object Classification & 1 human 1 AI & Sequential & Functional & Designated & Chain & Colocated & Long Term \\
13 & \cite{hemmer2025complementarity} & Classification & Object Classification & 1 human 1 AI & Sequential & Functional & Designated & Chain & Colocated & Long Term \\
13 & \cite{jia2021towards} & Classification & Object Classification & 1 human 1 AI & Sequential & Functional & Designated & Chain & Colocated & Long Term \\
13 & \cite{olla2024cybersecurity} & Cybersecurity/Security & Security & 1 human 1 AI & Sequential & Functional & Designated & Chain & Colocated & Long Term \\
13 & \cite{babbar2022utility} & Classification & Object Classification & 1 human 1 AI & Sequential & Functional & Designated & Chain & Colocated & Ad hoc \\
13 & \cite{mahmood2024designing} & Classification & Object Classification & 1 human 1 AI & Sequential & Functional & Designated & Chain & Colocated & Long Term \\
13 & \cite{bansal2021does} & Classification & Object Classification & 1 human 1 AI & Sequential & Functional & Designated & Chain & Colocated & Long Term \\
13 & \cite{zhang2022you} & Classification & Object Classification & 1 human 1 AI & Sequential & Functional & Designated & Chain & Colocated & Long Term \\
13 & \cite{josephs2023artifact} & Cybersecurity/Security & Object Classification & 1 human 1 AI & Sequential & Functional & Designated & Chain & Colocated & Long Term \\
13 & \cite{bansal2019case} & Experimental Game & Object Classification & 1 human 1 AI & Sequential & Functional & Designated & Chain & Colocated & Long Term \\
13 & \cite{bansal2019updates} & Experimental Game & Object Classification & 1 human 1 AI & Sequential & Functional & Designated & Chain & Colocated & Long Term \\
13 & \cite{smith2025navigating} & Cybersecurity/Security & Security & 1 human 1 AI & Sequential & Functional & Designated & Chain & Colocated & Long Term \\
13 & \cite{hong2023visualizing} & Education & Workbased Activity & 1 human 1 AI & Sequential & Functional & Designated & Chain & Colocated & Long Term \\
13 & \cite{zhao2025role} & Disaster Response & Search and Rescue & 1 human 1 AI & Sequential & Functional & Designated & Chain & Distributed & Ad hoc \\ \midrule
14 & \cite{jorge2024should} & Shopping & Teamwork Games & 1 human 2 AI & Sequential & Functional & Designated & Hub-and-Wheel & Colocated & Long Term \\ \midrule
15 & \cite{ye2022modeling} & Misc. Teamwork & Quiz & 4 humans 1 AI & Sequential & Functional & Distributed & Star & Colocated & Ad hoc \\ \bottomrule
\end{tabular}
}
\caption{All clusters of human-AI teams found in the analysis.}
\end{table}

\end{document}